\documentclass[twocolumn, 10pt, aps, superscriptaddress, floatfix, showpacs, prl, citeautoscript]{revtex4-1}

\usepackage[utf8]{inputenc}
\usepackage[T1]{fontenc}
\usepackage{graphicx}
\usepackage{amsmath}
\usepackage{amssymb}
\usepackage{wasysym}
\usepackage{upgreek}
\usepackage{bm}
\usepackage{xcolor}
\usepackage[colorlinks, citecolor={blue!50!black}, urlcolor={blue!50!black}, linkcolor={red!50!black}]{hyperref}
\usepackage{bookmark}
\usepackage{tabularx}
\usepackage{mathtools}
\usepackage{microtype}
\usepackage{bbm}
\usepackage[load=physical,load=abbr]{siunitx}

\newcommand{\co}[2]{#2}

\DeclarePairedDelimiter\abs{\lvert}{\rvert}%
\DeclarePairedDelimiter\norm{\lVert}{\rVert}%
\makeatletter
\let\oldabs\abs
\def\abs{\@ifstar{\oldabs}{\oldabs*}}
\let\oldnorm\norm
\def\norm{\@ifstar{\oldnorm}{\oldnorm*}}
\makeatother

\newcolumntype{L}[1]{>{\raggedright\arraybackslash}p{#1}}
\newcolumntype{C}[1]{>{\centering\arraybackslash}p{#1}}
\newcolumntype{R}[1]{>{\raggedleft\arraybackslash}p{#1}}

\graphicspath{{figures/}}

\begin{document}
\title{Breakdown of the Law of Reflection at a Disordered Graphene Edge}

\author{E. Walter}
\email[Electronic address: ]{elias.walter@rwth-aachen.de}
\affiliation{JARA Institute for Quantum Information, RWTH Aachen University,
 52056 Aachen, Germany}
\affiliation{Arnold Sommerfeld Center for Theoretical Physics,
Ludwig-Maximilians-University Munich, 80333 Munich, Germany}
\author{T. \"O. Rosdahl}
\email[Electronic address: ]{torosdahl@gmail.com}
\affiliation{Kavli Institute of Nanoscience, Delft University of Technology,
  P.O. Box 4056, 2600 GA Delft, Netherlands}
\author{A. R. Akhmerov}
\affiliation{Kavli Institute of Nanoscience, Delft University of Technology,
  P.O. Box 4056, 2600 GA Delft, Netherlands}
\author{F. Hassler}
\affiliation{JARA Institute for Quantum Information, RWTH Aachen University,
 52056 Aachen, Germany}

\date{August 24, 2018}

\begin{abstract}
  The law of reflection states that smooth surfaces reflect waves specularly, thereby acting as a mirror.
  This law is insensitive to disorder as long as its length scale is smaller than the wavelength.
  Monolayer graphene exhibits a linear dispersion at low energies and consequently a diverging Fermi wavelength.
  We present proof that for a disordered graphene boundary, resonant scattering off disordered edge modes results in diffusive electron reflection even when the electron wavelength is much longer than the disorder correlation length.
  Using numerical quantum transport simulations, we demonstrate that this phenomenon can be observed as a nonlocal conductance dip in a magnetic focusing experiment.
\end{abstract}

\maketitle
\emph{Introduction.}---The law of reflection is a basic physical phenomenon in
geometric optics.  As long as the surface of a mirror is flat on the scale of
the wavelength, a mirror reflects incoming waves specularly.  In the opposite
limit when the surface is rough, reflection is diffusive and an incident wave
scatters into a combination of many reflected waves with different angles.
This picture applies to all kinds of wave reflection, including sound waves
and particle waves in quantum systems. The phenomenon has been extensively
investigated both theoretically and experimentally in the past, e.g., in
order to understand sea clutter in radar \cite{davies1954} as well as a method to measure surface roughness \cite{bennett1961}.
\begin{figure}[!tb]
\includegraphics[width=0.97\columnwidth]{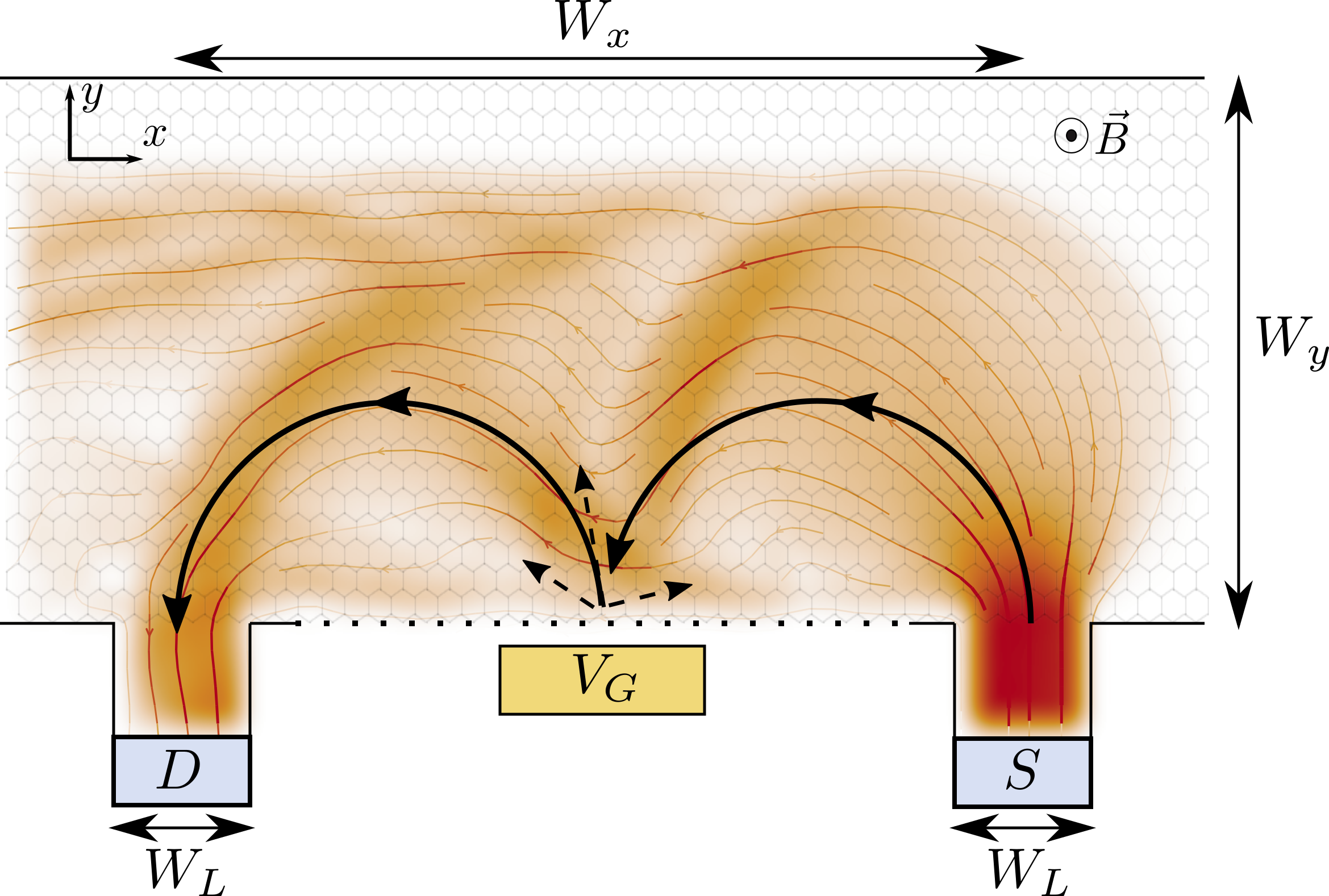}
\caption{Sketch of the setup.
Electrons injected at the source ($S$) follow cyclotron trajectories due to the perpendicular magnetic field $\boldsymbol{B} = B\hat{z}$, forming a hot spot at the boundary where most trajectories scatter.
If the trajectories specularly reflect at the boundary and the
separation $W_x$ between the midpoints of the source and the drain ($D$) matches
two cyclotron diameters, most trajectories enter the drain, and a focusing
peak manifests in the nonlocal conductance.
The focusing is evident in the classical cyclotron trajectory of an electron normally incident from $S$ at the Fermi level (solid curves),
and in the computed current distribution that is superimposed on the device (flow lines, colored background).
A side gate $V_G$ controls the average potential at the disordered boundary
(dotted line), and allows us to tune between regimes of specular and diffusive
reflection (see main text).
In the diffusive regime, electrons scatter into random angles as shown schematically with the dashed lines, resulting in a drop in the focusing peak conductance compared to the regime of specular reflection.
The graphene sheet is grounded, such that current due to off-resonance trajectories may drain away to the sides (open boundaries). }
\label{fig:sketch}
\end{figure}

\co{Graphene allows for optics experiments with electrons.}
Graphene \cite{geim2007,castro2009} is a gapless semiconductor with a linear dispersion relation near the charge neutrality point, and therefore a diverging Fermi wavelength.
Modern techniques allow for the creation of graphene monolayers of high mobility, with mean free paths of tens of microns \cite{lui2009, dean2010, mayorov2011, banszerus2016}.
This makes it possible to realize devices in which carriers propagate
ballistically over mesoscopic distances, facilitating the design of electron
optics experiments \cite{houten1990, cheianov2007, chen2016}.
For example, recent experiments employ perpendicular magnetic fields to demonstrate snaking trajectories in graphene \emph{p}-\emph{n} junctions \cite{rickhaus2015, taychatanapat2015}, or the magnetic focusing of carriers through cyclotron motion \cite{taychatanapat2013}.
The latter tests the classical skipping orbit picture of carrier propagation along a boundary \cite{bhandari2016}, and using a collimator to focus a narrow beam of electrons with a small angular spread enhances the focusing resolution \cite{barnard2017}.
The high mobility in the bulk together with a large Fermi wavelength suggest that graphene is a promising medium for the design of advanced electron optics and testing the law of reflection, cf.\ Fig.~\ref{fig:sketch}.

\co{We expect the law of reflection to work in graphene.}
Graphene edges are rough due to imperfect lattice termination or hydrogen passivation of dangling bonds \cite{son2006,halbertal2017}.
Boundary roughness may adversely affect device performance \cite{areshkin2007, evaldsson2008, masubuchi2012, dugaev2013}.
On the other hand, close to the charge neutrality point the Fermi wavelength in graphene diverges, and by analogy with optics, one may expect that the law of reflection holds and suppresses the diffusive  boundary scattering.

\co{Resonant edge mode scattering in graphene result in diffuse scattering, contrary to the law of reflection.}
In this Letter, we study how the microscopic boundary properties influence electron reflection off a graphene boundary.
Most boundaries result in the self-averaging of the boundary disorder, and therefore obey the law of reflection.
However, we find that, due to resonant scattering, electrons are reflected diffusively regardless of the Fermi wavelength when the disorder-broadened edge states overlap with $E=0$.
As a result, in this situation, the boundary of graphene never acts as a mirror and thus breaks the law of reflection.
We demonstrate that this phenomenon can be observed as a dip in the nonlocal conductance in a magnetic focusing setup (see Fig.~\ref{fig:sketch}).
We confirm our predictions by numerical simulations.

\co{We study scattering from a disordered zigzag boundary using the Dirac Hamiltonian, where we introduce disorder by randomly sampling the most general boundary condition for the Dirac Hamiltonian over the edge.}
\emph{Reflection at a disordered boundary.}---To demonstrate the breakdown of the law of reflection, we first analyze scattering at the edge of a semi-infinite graphene sheet.
We consider a zigzag edge, since the zigzag boundary condition applies to generic lattice terminations \cite{akhmerov2008}.
To begin with, we neglect intervalley scattering to simplify the analytical derivation, and focus on the single valley Dirac Hamiltonian
\begin{align}
H=v_F \, \boldsymbol{\sigma} \cdot\boldsymbol{p},
\end{align}
with $v_F$ the Fermi velocity, $\boldsymbol{\sigma}=(\sigma_x,\sigma_y)^T$ the vector of Pauli matrices in the (sublattice) pseudospin space, and $\boldsymbol{p}$ the momentum.
We later verify the validity of our conclusions with tight-binding calculations that include intervalley scattering.
We introduce edge disorder by randomly sampling the most general single-valley boundary condition \cite{berry1987, mccann2004, akhmerov2008} over the edge, such that the boundary condition for the wave function reads
\begin{align}
\psi(x,y=0) = [\cos\theta(x)\sigma_z + \sin\theta(x)\sigma_x]\, \psi(x,y=0), \label{eq:general_bc}
\end{align}
where disorder enters through the position-dependent parameter $\theta$, and $\theta = 0$ gives a zigzag segment.
We take $\theta(x)$ to follow a Gaussian distribution with mean value $\mathrm{E}[\theta(x)] = \theta_0$ and covariance $\mathrm{Cov}[\theta(x),\theta(x^\prime)] = s_\theta^2 e^{-\pi(x-x^\prime)^2/d^2}$, with $d$ the correlation length.
In this work, $\mathrm{E}[ A ]$ is the statistical average of $A$ over the disordered boundary, and the corresponding variance $\mathrm{Var}(A)$.
The boundary condition \eqref{eq:general_bc} applies to different microscopic origins of disorder, such as hydrogen passivation of dangling bonds \citep{akhmerov2008} or edge reconstruction \cite{vanostaay2011}.

\co{We define a scattering problem and the reflection angle as the average angle the incident modes scatter into, and use the variance of the reflection angle to quantify diffusiveness of reflection.}
To solve the scattering problem, we introduce periodic boundary conditions parallel to the boundary with period $L$, such that the momentum $k_\parallel \in \{2\pi n/L ~|~n \in \mathbb{Z}\}$ is conserved.
At the Fermi energy $E_F$, the disordered boundary scatters an incident mode $\psi^\mathrm{in}_{k_\parallel}$ into the outgoing modes $\psi^\mathrm{out}_{k^\prime_\parallel}$.
The scattering state is
\begin{align}
\psi_{k_\parallel} = \psi^\mathrm{in}_{k_\parallel} + \sum_{k^\prime_\parallel} \psi^\mathrm{out}_{k^\prime_\parallel} S_{k^\prime_\parallel  k_\parallel},
\label{eq:scattering_states}
\end{align}
where modes with $k_\parallel > k_F$ are evanescent but others propagating, with $k_F$ the Fermi momentum, and $S_{k^\prime_\parallel  k_\parallel}$ the reflection amplitudes.
An outgoing propagating mode moves away from the edge at the angle $\varphi_{k_\parallel} = \arctan(v_\parallel/v_\perp)$ relative to the boundary normal, with $v_\parallel$ and $v_\perp$ the velocities along and perpendicular to the boundary.
For the incident propagating mode at $k_\parallel$, the quantum mechanical average reflection angle is therefore
\begin{equation}
\langle \varphi_{k_\parallel} \rangle = \sum\limits_{k^\prime_\parallel} \varphi_{k^\prime_\parallel} |S_{k^\prime_\parallel k_\parallel}|^2,
\end{equation}
where the sum is limited to propagating modes, and $|S_{k^\prime_\parallel k_\parallel}|^2$ is the reflection probability into the outgoing mode at $k^\prime_\parallel$.
An incident mode reflects specularly if $S_{k^\prime_\parallel k_\parallel} = \delta_{k^\prime_\parallel k_\parallel}$, but diffusively if it scatters into multiple angles, and the variance $\sigma^2(\varphi_{k_\parallel})$ is therefore finite for the latter.
If $N$ modes are incident, diffusiveness manifests in a finite mode-averaged variance $\sigma^2(\varphi) = \sum_{k_\parallel} \sigma^2(\varphi_{k_\parallel})/N$, or its statistical average $\mathrm{E}[ \sigma^2(\varphi) ]$ over the disordered boundary.
If $\lambda_F \ll L$, then $\sigma^2(\varphi)$ automatically includes the statistical average $\mathrm{E}[\sigma^2(\varphi)]$, because the incident waves sample multiple different segments of the boundary within each period.

\co{The scattering problem simplifies at charge neutrality, where diffusiveness appears as a finite variance in the scattering phase.}
The scattering problem simplifies at the charge neutrality point $E_F=0$, where only two propagating modes are active, one incident and one outgoing, both with $k_\parallel = 0$.
The scattering matrix relating the propagating modes is therefore a phase factor $e^{i \phi}$, with $\phi$ the scattering phase, and the quantum mechanical averages of the preceding paragraph are not necessary.
We expect diffusiveness to manifest as a finite variance $\mathrm{Var}(\phi)$, and have verified this numerically.
To compute $\phi$, we impose the boundary condition \eqref{eq:general_bc} on the scattering state \eqref{eq:scattering_states}.

\co{If the disordered boundary condition has a finite mean, the variance of the scattering phase increases with boundary disorder, but decreases with boundary length, such that the boundary reflects specularly for extended boundaries.}
If $\theta_0$ is nonzero and $s_\theta \ll \theta_0$, $\phi$ follows a Gaussian distribution \cite{suppl} with the mean
\begin{align}
\mathrm{E}[\phi]
\overset{L\gg d}{=}& -\theta_0+\frac{s_\theta^2}{2\sin(\theta_0)}
+ \mathcal{O}\left(\frac{s_\theta^3}{\theta_0^3}\right)
\label{eq:mean_phi}
\end{align}
and variance
\begin{align}
\mathrm{Var}(\phi)
=&\frac{d}{L}s_\theta^2
+ \mathcal{O}\left(\frac{s_\theta^3}{\theta_0^3}\right)\,.
\label{eq:sigma_phi}
\end{align}
Thus $\mathrm{E}[\phi ]$ is given by $\theta_0$, with the addition of a random walklike drift term proportional to $s_\theta^2$.
In addition, $\mathrm{Var}(\phi)$ increases with $s_\theta^2$, but increasing the boundary length suppresses it as $1/L$.
In the limit $L\rightarrow\infty$ reflection is thus completely specular, with a fixed scattering phase $\phi$.
This algebraic decay of diffusive scattering resembles a classical optical mirror \cite{bennett1961}.

\co{If the mean of the disordered boundary condition vanishes, the variance of the scattering phase increases with boundary disorder, and is independent of boundary length, such that reflection does not become specular for an extended boundary.}
If $\theta_0=0$, surprisingly there is no suppression of $\mathrm{Var}(\phi)$ with $L$.
Rather, we find \cite{suppl} that $\tan\phi$ follows a Cauchy distribution $f(\tan{\phi})=\gamma/\pi(\tan^2\phi+\gamma^2)$
with $\mathrm{E} [\phi] = 0$, $\mathrm{Var}(\phi)\approx 2.2\,s_\theta$ linear in $s_\theta$ instead of quadratic, and $\gamma\approx 0.8 \, s_\theta$ obtained numerically.
In this case, the law of reflection therefore breaks down and scattering is always diffusive.
The distribution of the scattering phase follows the Cauchy distribution also when the disorder is non-Gaussian and even asymmetric, as long as $\theta_0$ is sufficiently small.
For an asymmetric distribution, the value of $\gamma/s_\theta$ weakly depends on higher cumulants of the distribution of $\theta(x)$.

\begin{figure}[!tb]
\includegraphics[width=0.97\columnwidth]{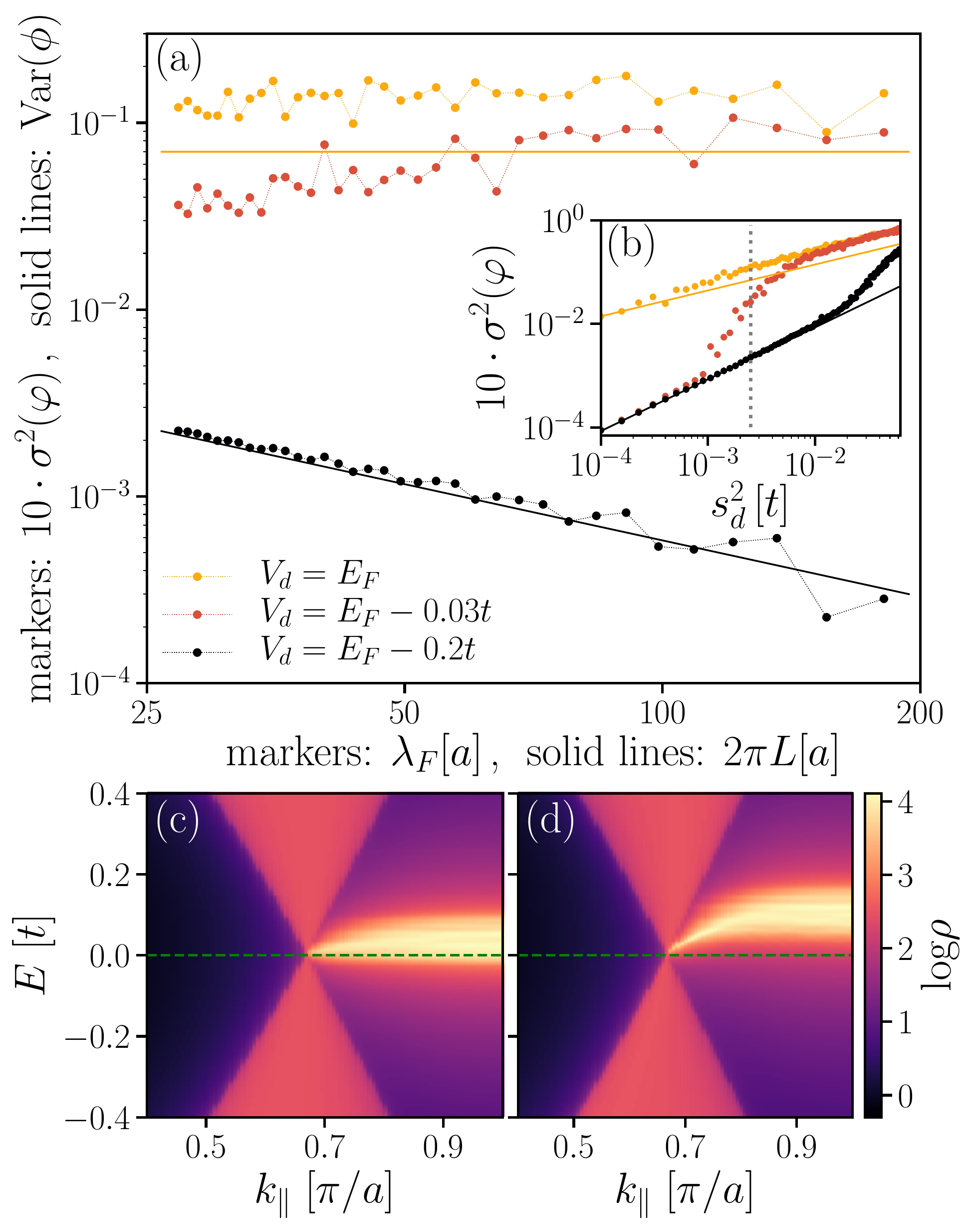}
\caption{(a) Solid lines: $\text{Var}(\phi)$ at the Dirac
  points ($E_F=0$) as a function of the boundary length $L$,
  for a disorder strength $s_d=0.05t$ obtained from the tight-binding model.
Markers: $\sigma^2(\varphi)$ at finite $E_F$, averaged over all incoming modes and $10^2$ disorder configurations, as a function of the Fermi wavelength $\lambda_F$ for the same disorder strength, obtained numerically for a semi-infinite graphene sheet with a boundary of length $L=300a$.
The values chosen for $\lambda_F=\sqrt{3}\pi ta/E_F$ correspond to $E_F$ ranging from $0.2t$ to $0.03t$.
(b) Same as (a), as a function of the disorder strength $s_d^2$, for a value of $2\pi L \approx 27 a$ [$\lambda_F \approx 27 a$, $E_F = 0.2 t$]. The dotted line indicates the value of $s_d$ used in (a). 
For $V_d=E_F$ the variances of both the scattering phase at $E_F=0$ and the reflection angle at $E_F>0$ increase linearly with $s_d$, independent of the Fermi wavelength, exhibiting the breakdown of the law of reflection.
For $|V_d - E_F| \gtrsim s_d$, $\mathrm{Var}(\phi)$ [$\sigma^2(\varphi)$] decays with increasing $L$ [$\lambda_F$] as $1/L$ [$1/\lambda_F$] and increases quadratically with the disorder strength [as given by Eq.~\eqref{eq:sigma_phi}]. Reflection is thus specular, but becomes diffusive for $|V_d - E_F| \lesssim s_d$.
Setting $V_d$ closer to $E_F$ moves transition between the regimes of specular and diffusive reflection to smaller $s_d$.
This is because of the overlap of $E_F$ with the disorder-broadened edge band.
(c),(d) Momentum-resolved density of states at the disordered zigzag edge of a semi-infinite graphene sheet with a boundary of length $L = 300a$.
A band of edge states with bandwidth $\propto s_d = 0.05t$ extends between the Dirac cones, residing mostly at energy $V_d$, with $V_d = 0.03t$ in (c) and $V_d = 0.2t$ in (d) [dashed lines].}
\label{fig:sigma_phi}
\end{figure}

\co{The breakdown of the law of reflection happens because of electron scattering off disordered edge modes.}
Generic graphene boundaries support bands of edge states with a linear dispersion \cite{akhmerov2008, vanostaay2011}.
Because the matrix element between the edge state and the edge disorder is inversely proportional to the spatial extent of the edge state, the disorder broadening of these edge states is proportional to the momentum along the boundary [see Figs.~\ref{fig:sigma_phi}(c),~\ref{fig:sigma_phi}(d)].
In other words linearly dispersing edge states turn into disorder-broadened bands with both the average velocity and the bandwidth proportional to $k_\parallel$.
When these bands overlap with $E=0$ they serve as a source of resonant scattering responsible for the breakdown of the law of reflection.
Indeed, we find that the condition for diffusive scattering occurs for any $\theta_0 \lesssim s_\theta$.

\co{We repeat the preceding analysis at the Dirac point using the tight-binding model, in order to include both inequivalent valleys,
and find identical results up to a prefactor.}
To include intervalley scattering, we compute the scattering phase at the charge neutrality point using the nearest neighbor tight-binding model of graphene, with random on-site disorder in the outermost row of atoms taken from a Gaussian distribution with mean $V_d$ and variance $s_d^2$ \cite{suppl}.
The results, shown in Fig.~\ref{fig:sigma_phi}(b), agree with the single valley prediction of the Dirac equation up to numerical prefactors.

\co{At finite Fermi energy, we solve the scattering problem numerically and find that in the presence of edge disorder, the boundary reflection angle behaves in qualitative agreement with preceding results for the scattering phase at the Dirac points.}
To extend our analysis to nonzero $E_F$, we employ the tight-binding model with on-site disorder to study the reflection angle $\varphi$ at the disordered boundary numerically using Kwant \cite{groth2014}.
The disordered edge band now resides at the energy $V_d$, as Figs.~\ref{fig:sigma_phi}(c) and \ref{fig:sigma_phi}(d) show.
Figures~\ref{fig:sigma_phi}(a), \ref{fig:sigma_phi}(b) confirm that $\sigma^2(\varphi) \approx \mathrm{Var}(\phi)$ at $E=0$.
The law of reflection is broken for all $s_d$ at $V_d=E_F$ and $\mathrm{Var}(\phi)$ increases linearly with $s_d$, independent of $\lambda_F$.
Further, the reflection becomes specular for $s_d \lesssim |V_d - E_F|$.
As Fig.~\ref{fig:sigma_phi}(b) shows, $\mathrm{Var}(\phi)$ [$\sigma^2(\varphi)$] increases quadratically with the disorder strength $s_d$, but decays as $1/L$ [$1/\lambda_F$] (Fig.~\ref{fig:sigma_phi}(a)) when the Fermi wavelength becomes large compared to the lattice constant $a$, such that scattering is predominantly specular.
However, for $s_d \gtrsim |V_d - E_F|$ reflection becomes diffusive, and moving $V_d$ closer to $E_F$ [Fig.~\ref{fig:sigma_phi}(b)] shifts the transition from specular to diffusive reflection to smaller $s_d$.

\emph{Experimental detection.}---\co{We propose a transport measurement based on magnetic focusing to probe the disordered boundary.}
Any experiment that is sensitive to the microscopic properties of a disordered boundary will detect the breakdown of the law of reflection if the disordered edge band overlaps with the Fermi level.
We propose to search for a transport signature of the breakdown of the law of reflection in the magnetic focusing experiment sketched in Fig.~\ref{fig:sketch}.
The idea is to study the reflection of ballistic cyclotron trajectories in a
magnetic field $B$ off a graphene edge \cite{houten1990, taychatanapat2013, bhandari2016}.
The use of a collimator could improve such an experiment \cite{barnard2017}.
\begin{figure}[!tb]
\includegraphics[width=0.97\columnwidth]{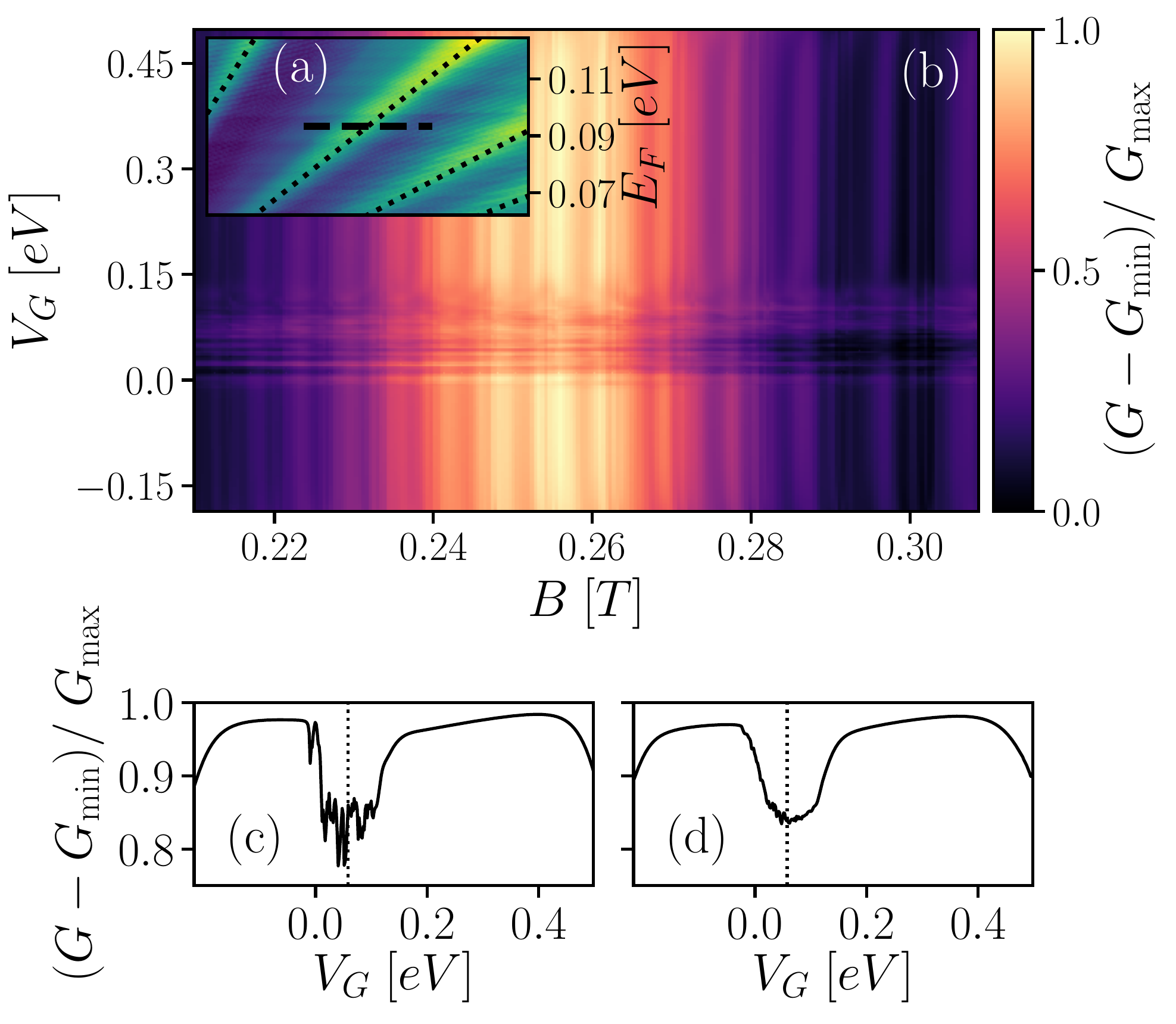}
\caption{(a) Conductance as a function of Fermi energy and magnetic field showing the first $4$ magnetic focusing peaks for the device sketched in Fig.~\ref{fig:sketch} in the absence of edge disorder and with $V_G = 0$.
Superimposed are the predicted locations of the focusing peaks (dotted lines), $1 \leq p \leq 4$ from left to right across the diagonal.
The color scale is linear and ranges from about $4e^2/h$ (dark) to $28e^2/h$ (bright).
(b) Conductance around the $p=2$ focusing peak at $E_F = 0.093$ eV [dashed line in (a)] versus gate voltage.
We include disorder with $V_d = 0.062$ eV and $s_d = 0.047$ eV in the
first $N=6$ rows next to the boundary.
Reflection at the boundary is specular and the conductance smooth in $V_G$, except for a dip when the disordered edge band overlaps with the Fermi level, and reflection becomes diffusive.
(c) Line cut from (b) at $B=0.256\,$T with the predicted voltage value for the dip marked.
Within the dip, the conductance exhibits fluctuations dependent on the particular disorder configuration, that are washed out by disorder averaging in (d).
We assume the scaling factor $s = 9$ in the tight-binding model, such that $W_x = 1.6 \ \upmu\mathrm{m}$, $W_y=1 \ \upmu\mathrm{m}$ and $W_L = 0.2 \ \upmu\mathrm{m}$.}
\label{fig:focusing}
\end{figure}

\co{Magnetic focusing happens when the separation between source and drain is an integer multiple of the cyclotron diameter.
We expect a reduction in focusing conductance if the boundary scattering is diffusive, in comparison with a boundary that is specularly reflective.}
Magnetic focusing refers to the appearance of peaks in the nonlocal conductance between the source and the drain when a voltage is applied between the source and the grounded ribbon, cf.\ Fig.~\ref{fig:sketch}.
There is an increased probability for electrons to end up in the drain whenever the separation $W_x$ between source and drain matches an integer multiple of
the cyclotron diameter $2r_c$, where $r_c = \hbar k_F/eB$ is the cyclotron radius with $k_F$ the Fermi momentum, $\hbar$ the reduced Planck constant, and $e$ the elementary charge.
Due to the linear dispersion near the charge neutrality point in graphene, $k_F = E_F/\hbar v_F$ is linear in $E_F$, such that focusing peaks appear at the magnetic fields $B^f_n = 2n E_F/ev_F W_x$, $n \in \mathbb{N}$.
For the setup in Fig.~\ref{fig:sketch} but with a clean, specularly reflecting system edge, Fig.~\ref{fig:focusing}(a) shows a map of the first few focusing conductance peaks with their predicted locations marked.
At resonance $p$, the electron beam reflects specularly $p-1$ times at the system edge before exiting into the drain, as Fig.~\ref{fig:sketch} demonstrates for $p=2$.
On the other hand, if reflection from the boundary is diffusive, the electrons scatter into random angles off the boundary, which in general no longer result in cyclotron trajectories that are commensurate with the distance from the focus point at the boundary to the drain.
In comparison with the case of specular reflection, the focusing beam at the drain is therefore diminished for diffusive edge scattering, resulting in a drop in the $p>1$ conductance resonances.
Because the reflection is diffusive when the disordered edge band overlaps with the Fermi level, by using a side gate (see Fig.~\ref{fig:sketch}) to tune the average potential at the disordered boundary, it is therefore possible to observe signatures of the breakdown of the law of reflection in the form of a conductance drop at a focusing peak.

\co{We simulate the device with a tight-binding model using Kwant, and our
results apply to realistic systems by scaling of the tight-binding model.}
To verify our prediction, we perform numerical simulations of the graphene
focusing device with a side gate sketched in Fig.~\ref{fig:sketch}.
We implement the tight-binding model for graphene in Kwant \cite{groth2014} and include the magnetic field via a Peierls substitution.
We apply a random uniformly distributed onsite potential with mean $V_d$ and variance $s_d^2$ to the first several rows of atoms adjacent to the system edge.
We simulate the effect of a side gate by applying an extra potential with amplitude $V_G$ exponentially decaying away from the sample edge on a length scale comparable to
the size of the disordered region.
Away from the charge neutrality point, we expect peak diffusive edge scattering to occur when the average potential by the boundary matches the Fermi energy.
The relevant scales for our simulations are the hopping $t$, the graphene lattice constant $a=2.46\,$\AA, and the
magnetic flux $\Phi \propto B a^2$ per unit cell.
Scaling the tight-binding Hamiltonian with a scaling factor $s$ \cite{liu2015} by reinterpreting $t/s \equiv t$, $sa \equiv a$ and $B/s^2 \equiv B$ such that $\Phi$ is unchanged by the scaling, our simulations apply to graphene devices of realistic and experimentally realizable dimensions \cite{taychatanapat2013, bhandari2016}.
Note that the onsite disorder correlation length is not scale invariant, and the disorder thus correlates $s$ lattice sites in the original model.

\co{We observe a drop in the second conductance resonance when the side gate is used to tune the average boundary potential such that the reflection is diffusive.}
Tuning the average potential at the disordered system edge by varying the side gate $V_G$ reveals a clear dip in the conductance Fig.~\ref{fig:focusing}(b) around the second focusing resonance $p=2$, which is absent when no edge disorder is included \cite{suppl}.
Outside the dip the conductance only changes weakly with $V_G$, which is the expected behavior for a clean specularly reflecting boundary.
Here, the first $N=6$ rows of sites adjacent to the edge are disordered, and the extent of the disordered region into the graphene sheet thus approximately $2.1 a \ll \lambda_F \approx 18a$, such that the length scales are consistent with specular reflection.
The conductance fluctuates erratically within the dip, as the line cut Fig.~\ref{fig:focusing}(c) taken from Fig.~\ref{fig:focusing}(b) at $B=0.256\,$T shows.
These are universal conductance oscillations particular to an individual disorder configuration.
They are washed out by disorder averaging as Fig.~\ref{fig:focusing}(d) shows, revealing an omnipresent conductance dip.
Furthermore, the conductance dip appears when the disordered edge band overlaps with $E_F$, which is the condition for the breakdown of the law of reflection, with the $V_G$ that aligns the band with $E_F$ marked in Figs.~\ref{fig:focusing}(c) and \ref{fig:focusing}(d).

\emph{Conclusion and discussion.}---Our analysis of scattering at a disordered graphene boundary reveals a regime where specular reflection is suppressed in favor of diffusive scattering.
This counterintuitive conclusion holds even when conventional wisdom dictates that specular reflection should dominate and the boundary should act as a mirror, namely when a boundary is rough on a length scale smaller than the Fermi wavelength.
The origin of this breakdown of the law of reflection is resonant scattering of the electron waves from a linear superposition of localized boundary states.
Our calculations show that this phenomenon is detectable in transverse magnetic focusing experiments, by employing a side gate to tune the average potential at the boundary.
In these experiments the breakdown of specular reflection manifests as a dip in the nonlocal conductance at the second focusing resonance.
Because the zigzag boundary condition is generic in graphene, we expect our results to apply to an arbitrary termination direction, and to be insensitive to microscopic details.
We are thus confident that this effect is experimentally observable in present-day devices.

\acknowledgments
This work was supported by ERC Starting Grant No.~638760, the Netherlands Organisation for Scientific Research (NWO/OCW), and the U.S. Office of Naval Research.

\bibliographystyle{apsrev4-1}
\bibliography{refs}


\onecolumngrid\clearpage
\setcounter{equation}{0}
\renewcommand\theequation{S\arabic{equation}}
\setcounter{figure}{0}
\renewcommand\thefigure{S\arabic{figure}}
\section{Supplement}\setcounter{page}{1}

\subsection{Computation of the scattering phase in the continuum description}

In the following, we present the derivation of the scattering phase at $E_F=0$ from the continuum description governed by the Dirac equation, which is valid within the linear regime of the graphene dispersion.

\begin{figure}[!tbh]
\centering
\includegraphics[width=.48\columnwidth]{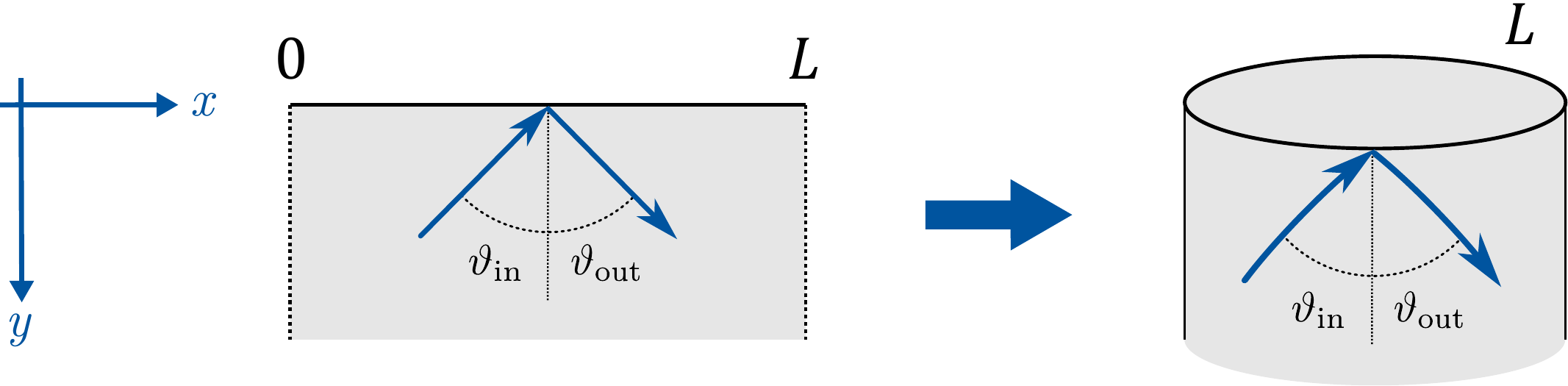}
\caption{Scheme of the system geometry: A graphene sheet (gray) with translational invariance in $y$-direction is terminated by a single boundary at $y=0$. Applying periodic boundary conditions (left, dotted lines) in $x$-direction on the semi-infinite plane is equivalent to rolling it up to a cylinder (right). $L$ is the boundary length after applying periodic boundary conditions. Blue arrows indicate schematically the paths of an incoming and an outgoing mode, with angles relative to the surface normal of $\vartheta_\mathrm{in}$ and $\vartheta_\mathrm{out}$, respectively.}
\label{fig:geometry}
\end{figure}
We consider a cylindrical geometry as sketched in Fig.~\ref{fig:geometry} with a boundary of length $L$, which in the limit $L\rightarrow\infty$ resembles a semi-infinite sheet with a single boundary at $y=0$. We describe electronic properties in terms of the Dirac Hamiltonian of a single valley,
\begin{align}
H=v_F\, \boldsymbol{\sigma}\cdot\mathbf{p}
=-i\hbar v_F\left(
\begin{matrix}
0 & \partial_x-i\partial_y \\ \partial_x+i\partial_y & 0
\end{matrix}
\right)\,,
\end{align}
as defined in the main text.
With the ansatz
$
\psi(\mathbf{r}) = e^{i\mathbf{q}\cdot\mathbf{r}}
\left(\psi_A, \psi_B\right)^T
$
we obtain from the Dirac equation at zero energy $H\psi=0$
\begin{align}
\left\lbrace
\begin{array}{r}
(q_x-iq_y)\psi_B = 0 \,, \\
(q_x+iq_y)\psi_A = 0 \,.
\end{array}
\right.
\label{eq:Dirac_eq}
\end{align}
Periodic boundary conditions in $x$-direction $\psi(x,y)=\psi(x+L,y)$ restrict the momentum $q_x=2\pi n/L$, with $n\in \mathbb{Z}$.
With the boundary at $y=0$ and the graphene sheet extending to positive $y$ as shown in Fig.~\ref{fig:geometry}, we can write down all non-trivial solutions of Eq.~\eqref{eq:Dirac_eq} for given $n$.
We can distinguish two cases, depending on the behavior for $y\rightarrow \infty$:

For $n=0$ we have $\mathbf{q}=0$ and therefore all states $\psi=(\psi_A,\psi_B)^T$ are solutions to the Dirac equation \eqref{eq:Dirac_eq}. We can choose an orthonormal basis $\{\psi_+,\psi_-\}$ of that two-dimensional subspace that diagonalizes the $y$-component of the current operator $\mathbf{J}=v_F \boldsymbol{\sigma}$, such that $\psi_\pm$ have well-defined current $\pm v_F$ perpendicular to the boundary,
\begin{align}
\psi_\eta^\dagger J_y \psi_\nu &= \eta v \delta_{\eta \nu}\,,\quad \eta,\nu=\pm\,,
\\
\psi_\eta^\dagger \psi_\nu &= \delta_{\eta\nu}\,.
\end{align}
The propagating modes are therefore the eigenstates of $\sigma_y$ that can be written as
$
\psi_\pm = \frac{1}{\sqrt{2}}(1,\pm i)^T
$.
As $\psi_-$ has a velocity $-v_F$ and is thus moving in negative $y$-direction, we consider it to be incoming and $\psi_+$ to be outgoing, respectively.

For $n\neq 0$ the Dirac equation \eqref{eq:Dirac_eq} becomes
\begin{align}
\left\lbrace
\begin{array}{r}
\left(2\pi n/L -iq_y\right) \psi_B = 0\,,
\\
\left(2\pi n/L +iq_y\right) \psi_A = 0\,.
\end{array}
\right.
\end{align}
such that we get two non-trivial solutions for each $n$:
For $q_y=-2\pi i n/L$ and $\psi_A=0$ we have
$\psi_{n,-}=e^{2\pi i n x/L}e^{2\pi n y/L}
(0,1)^T$.
This mode decays exponentially into the bulk for $y\rightarrow \infty$ if $n<0$, but is not normalizable for positive $y$ if $n>0$.
For $q_y=2\pi i n/L$ and $\psi_B=0$ we have
$\psi_{n,+}=e^{2\pi i n x/L}e^{-2\pi n y/L}
(1,0)^T$.
This mode is evanescent if $n>0$, but not normalizable if $n<0$.
In total we thus remain with one evanescent mode for each $n\in\mathbb{Z}\setminus \{0\}$.

We can now construct a scattering state $\psi$ from the incoming mode $\psi_-$, outgoing mode $\psi_+$ and evanescent modes $\psi_{n,\pm}$ as
\begin{align}
\psi &= \psi_- + S\psi_+ + \sum\limits_{n=1}^{\infty} (\alpha_n \psi_{n,+}+\alpha_{-n}\psi_{-n,-})\,,
\end{align}
where $S=e^{i\phi}$ is the scattering phase that the incoming mode acquires when scattered into the outgoing one, and $\alpha_n$ is the amplitude to scatter into the $n$-th evanescent mode.
A boundary is introduced by requiring this scattering state to fulfill the boundary condition
\begin{align}
M\psi(x,y=0)=\psi(x,y=0)\,.
\label{eq:general_boundary_condition}
\end{align}
A disordered boundary interpolating between a clean zigzag boundary and an infinite-mass (Berry-Mondragon \cite{berry1987}) boundary condition constitutes the most general single-valley boundary condition. This boundary condition applies to different microscopic origins of disorder, such as the staggered potential on a zigzag boundary which is produced by a passivation of the dangling bonds \cite{akhmerov2008}, or effects of edge reconstruction \cite{vanostaay2011}. 
The zigzag boundary is given by the matrix $M_{zz}=\sigma_z$, whereas the Berry-Mondragon boundary is specified by $M_\mathrm{BM}=\boldsymbol{\sigma}\cdot (\hat{\mathbf{z}}\times \mathbf{n}_B)=\sigma_x$ for the boundary normal $\mathbf{n}_B=-\hat{\mathbf{y}}$. We therefore consider the boundary condition matrix
\begin{align}
M = \cos\theta(x) M_{zz} + \sin\theta(x) M_\mathrm{BM}
=\left(
\begin{matrix}
\cos\theta(x) & \sin\theta(x) \\
\sin\theta(x) & -\cos\theta(x)
\end{matrix}
\right)\,,
\end{align}
with a random function $\theta(x)$ to introduce disorder by a spatially fluctuating staggered potential, such that we obtain a zigzag boundary for $\theta=0$ and an infinite-mass boundary for $\theta=\pi/2$. The value of $\theta(x)$ at the position $x$ on the boundary is randomly taken from a Gaussian distribution with mean value $\theta_0$ and variance $s_\theta^2$.
Furthermore, we assume a Gaussian correlation in space,
\begin{align}
\mathrm{Cov}[\theta(x),\theta(x^\prime)] = s_\theta^2 e^{-\pi(x-x^\prime)^2/d^2} \,,
\label{eq:theta_correlations}
\end{align}
with a correlation length $d$ that corresponds to a lattice constant, since the real problem lives on a lattice. In the limit $d\rightarrow 0$ the correlations become $\mathrm{Cov}[\theta(x),\theta(x^\prime)] \rightarrow s_\theta^2 d \delta(x-x^\prime)$.

With $\psi(x,y=0) = (\psi_A(x), \psi_B(x))^T$ we obtain from the boundary condition Eq.~\eqref{eq:general_boundary_condition}
\begin{align}
\mu(x)\sum_{n=0}^\infty \alpha_{n} e^{2\pi inx/L}
- \sum_{n=-1}^{-\infty} \alpha_n e^{2\pi inx/L}
-i \alpha_0
=  - i\sqrt{2} \,,
\label{eq:boundary_condition_equation_infinite}
\end{align}
with $\mu(x) = \tan(\theta(x)/2)$ (being 0 for a clean zigzag and 1 for the infinite-mass type boundary) and $\alpha_0=(1+S)/\sqrt{2}$.
We Fourier-transform Eq.~\eqref{eq:boundary_condition_equation_infinite} by applying to both sides $\tfrac{1}{L}\int_0^L \mathrm{d}x \, e^{-2\pi i mx/L}$, with $m\in \mathbb{Z}$, to obtain
\begin{align}
\sum_{n=1}^\infty \tilde{\mu}_{n-m} \alpha_n
+ (\tilde{\mu}_{-m} - i\delta_{m,0}) \alpha_0
- \sum_{n=-1}^{-\infty} \delta_{m,n}\alpha_n
= -i\sqrt{2} \, \delta_{m,0} \,,
\label{eq:boundary_condition_equation_Fourier}
\end{align}
with the Fourier components of the disorder function $\mu$,
\begin{align}
\tilde{\mu}_m = \frac{1}{L} \int\limits_0^L \mathrm{d}x \, e^{2\pi i m x/L} \mu(x)\,, \quad m\in\mathbb{Z}\,.
\end{align}
We can rephrase Eq.~\eqref{eq:boundary_condition_equation_Fourier} in matrix form as
\begin{align}
\underbrace{
\left(
\renewcommand{\arraystretch}{1.3}
\begin{array}{c c c}
\boldsymbol{\tilde{\mu}} & \tilde{\mu}_\uparrow & 0 \\
\tilde{\mu}_\uparrow^\dagger & \tilde{\mu}_0 - i & 0 \\
\boldsymbol{\tilde{\mu}^\prime} & \tilde{\mu}_\downarrow & -\mathbbm{1}
\end{array}
\right)
}_{\boldsymbol{\tilde{A}}}
\cdot
\left(
\renewcommand{\arraystretch}{1.3}
\begin{array}{c}
\alpha_+ \\ \alpha_0 \\ \alpha_-
\end{array}
\right)
=
\left(
\renewcommand{\arraystretch}{1.3}
\begin{array}{c}
0 \\ -i\sqrt{2} \\ 0
\end{array}
\right)\,,
\label{eq:matrix_equation_Fourier}
\end{align}
with
\begin{align}
\alpha_+ &=
\left(
\begin{array}{c}
\vdots \\ \alpha_3 \\ \alpha_2 \\ \alpha_1
\end{array}
\right)
\,, \quad
\alpha_- =
\left(
\begin{array}{c}
\alpha_{-1} \\ \alpha_{-2} \\ \alpha_{-3} \\ \vdots
\end{array}
\right)\,, \quad \text{and}
\nonumber \\
\boldsymbol{\tilde{\mu}} &=
\left(
\begin{array}{c c c c}
\ddots & \ddots & \ddots & \vdots \\
\ddots & \tilde{\mu}_0 & \tilde{\mu}_1^* & \tilde{\mu}_2^*\\
\ddots & \tilde{\mu}_1 & \tilde{\mu}_0 & \tilde{\mu}_1^* \\
\cdots & \tilde{\mu}_2 & \tilde{\mu}_1 & \tilde{\mu}_0
\end{array}
\right)
\,,\quad
\boldsymbol{\tilde{\mu}^\prime} =
\left(
\begin{array}{c c c c}
\cdots & \tilde{\mu}_4 & \tilde{\mu}_3 & \tilde{\mu}_2 \\
\text{\reflectbox{$\ddots$}} & \tilde{\mu}_5 & \tilde{\mu}_4 & \tilde{\mu}_3 \\
\text{\reflectbox{$\ddots$}} & \tilde{\mu}_6 & \tilde{\mu}_5 & \tilde{\mu}_4 \\
\text{\reflectbox{$\ddots$}} & \text{\reflectbox{$\ddots$}} & \text{\reflectbox{$\ddots$}} & \vdots
\end{array}
\right)\,,
\tilde{\mu}_\uparrow =
\left(
\begin{array}{c}
\vdots \\ \tilde{\mu}_3^* \\ \tilde{\mu}_2^* \\ \tilde{\mu}_1^*
\end{array}
\right)
\,, \quad
\tilde{\mu}_\downarrow =
\left(
\begin{array}{c}
\tilde{\mu}_1 \\ \tilde{\mu}_2 \\ \tilde{\mu}_3 \\ \vdots
\end{array}
\right) \,.
\label{eq:mu_matrices}
\end{align}
Hence, we have transformed the general boundary condition Eq.~\eqref{eq:general_boundary_condition} into a system of equations for the scattering phase (expressed through $\alpha_0$). This system is specified by the Fourier coefficients of the disorder function $\mu$. To solve Eq.~\eqref{eq:matrix_equation_Fourier} for $S$, we have to invert $\boldsymbol{\tilde{A}}$ to obtain
$
S = \sqrt{2}\alpha_0 - 1
=-1 -2i \,(\boldsymbol{\tilde{A}}^{-1})_{0,0}\,,
$
where $(\boldsymbol{\tilde{A}}^{-1})_{0,0}$ is the component in the center of $\boldsymbol{\tilde{A}}^{-1}$, referring to the $n=m=0$ Fourier components.

Due to the Gaussian correlation of $\theta(x)$ in space (Eq.~\eqref{eq:theta_correlations}), the Fourier components \mbox{$\tilde{\theta}_n=\tfrac{1}{L} \int_0^L \mathrm{d}x \, e^{2\pi i nx/L} \theta(x)$} decay for large $n$,
\begin{align}
\mathrm{E} [\tilde{\theta}_n ] = \theta_0 \delta_{n,0}\,, \quad
\mathrm{Cov} [\tilde{\theta}_n^*, \tilde{\theta}_m ]
\overset{d\ll L}{\approx} \delta_{n,m} \frac{s_\theta^2}{\sqrt{2\pi n_0^2}} \,e^{-n^2/2n_0^2} \,,
\label{eq:theta_tilde_correlations}
\end{align}
on a length scale $n_0 = L/\sqrt{2\pi}d$. The same holds for $\tilde{\mu}_n$, hence we can imagine to cut off at some $N\gg n_0$, such that the matrices in Eq.~\eqref{eq:mu_matrices} are finite-dimensional and we can safely use standard formulae for block-wise matrix inversion to formally obtain
\begin{align}
S = \frac{i-\tilde{m}}{i+\tilde{m}} \,,
\end{align}
with $\tilde{m} = \tilde{\mu}_\uparrow^\dagger \boldsymbol{\tilde{\mu}}^{-1} \tilde{\mu}_\uparrow - \tilde{\mu}_0$, and therefore
\begin{align}
\phi = \arg (S) = \text{atan2}\left(\mathfrak{Re}(S),\mathfrak{Im}(S)\right)
= \text{atan2}\left(1-\tilde{m}^2,2\tilde{m} \right) \,,
\label{eq:phi_Dirac}
\end{align}
where the atan2-function is closely related to the arctangent but adjusted such that it properly gives the angle between its arguments.

The inversion of $\boldsymbol{\tilde{\mu}}$ is not generically possible. However, an approximate solution can be found when $\boldsymbol{\tilde{\mu}}$ is dominated by its diagonal. We split up $\theta(x)$ into its mean value and fluctuations, $\theta(x)=\theta_0 + \delta\theta(x)$, with
\begin{align}
\mathrm{E}[\delta\theta(x)] &= 0\,, \\
\mathrm{Cov}[\delta\theta(x),\delta\theta(x^\prime)] &= s_\theta^2 e^{-\pi(x-x^\prime)^2/d^2} \,,
\label{eq:corr_delta_theta}
\end{align}
according to Eq.~\eqref{eq:theta_correlations}.
Assuming the disorder to be weak, $s_\theta\ll 1$, we can similarly expand $\mu(x)=\tan(\theta(x)/2)$ to get
\begin{align}
\mu(x) &= \tan\left(\frac{\theta_0}{2}\right) + \frac{1+\tan^2\left(\frac{\theta_0}{2}\right)}{2} \, \delta\theta(x) + \mathcal{O}\left(\delta\theta(x)^2\right)
\nonumber \\
&= \mu_0 + \delta\mu(x) + \mathcal{O}\left(\delta\mu(x)^2\right)\,.
\label{eq:mu_0}
\end{align}
The Fourier coefficients read
\begin{align}
\tilde{\mu}_n =\mu_0\,\delta_{n,0} + s_\mu \tilde{x}_n\,,
\label{eq:mu_n_tilde_expanded}
\end{align}
where
\begin{align}
s_\mu = \frac{1+\tan^2\left(\frac{\theta_0}{2}\right)}{2}\, s_\theta
\label{eq:sigma_mu}
\end{align}
is the standard deviation of $\mu$ and
\begin{align}
\tilde{x}_n = \frac{1}{L}\int\limits_0^L \mathrm{d}x \, e^{2\pi inx/L}\ \frac{\delta\mu(x)}{s_\mu}
\label{eq:x_n_tilde}
\end{align}
is normalized to have variance 1 and by definition a mean value of 0. Furthermore, from Eq.~\eqref{eq:theta_tilde_correlations} we see that
\begin{align}
\mathrm{Cov} (\tilde{x}_n^*, \tilde{x}_m ) \overset{d\ll L}{\approx} \delta_{n,m} \frac{1}{\sqrt{2\pi n_0^2}} \,e^{-n^2/2n_0^2} \,.
\label{eq:x_tilde_correlations}
\end{align}
With Eq.~\eqref{eq:mu_n_tilde_expanded} we get
\begin{align}
\boldsymbol{\tilde{\mu}} =
\mu_0\,\mathbbm{1}_N
+
s_\mu
\boldsymbol{\tilde{x}}\,,
\label{eq:mu_tilde}
\end{align}
thereby splitting it up into a diagonal part which is trivial to invert and a random Toeplitz matrix
\begin{align}
\boldsymbol{\tilde{x}}=
\left(
\begin{matrix}
\ddots & \ddots & \vdots \\
\ddots & \tilde{x}_0 & \tilde{x}_1^* \\
\cdots & \tilde{x}_1 & \tilde{x}_0
\end{matrix}
\right)\,,
\label{eq:x_tilde}
\end{align}
that cannot be inverted explicitly analytically.

For $\theta_0=0=\mu_0$, the disorder potential $\theta(x)$ is zero on average such that the disorder-broadened edge states overlap with $E=0$, whereas a finite $\theta_0>s_\theta$ (or $\mu_0>s_\mu$) shifts them away from $E=0$. We can directly translate these two cases to the structure of $\boldsymbol{\tilde{\mu}}$:
\begin{itemize}
\item For finite $\mu_0$ with small fluctuations $s_\mu$ on top, $\boldsymbol{\tilde{\mu}}$ is dominated by its diagonal. Hence, we can expand its inverse in powers of $s_\mu$. In this case, where the law of reflection is expected to hold, we can therefore give an explicit expression for $\phi$ for sufficiently weak disorder.
\item For a boundary with $\mu_0=0$ that fulfills the condition for diffusive scattering, this consideration does not work as then $\boldsymbol{\tilde{\mu}}=s_\mu \boldsymbol{\tilde{x}}$. In this case we have to rely on a numerical analysis.
\end{itemize}

\subsubsection{Scattering phase if law of reflection holds}

In the limit where $s_\mu\ll\mu_0$, we can expand
\begin{align}
\boldsymbol{\tilde{\mu}}^{-1}=\frac{1}{\mu_0}\ \mathbbm{1}-\frac{s_\mu}{\mu_0^2}\ \boldsymbol{\tilde{x}} + \mathcal{O}\left(\frac{s_\mu^2}{\mu_0^2}\right)
\label{eq:mu_tilde_inverse_approx}
\end{align}
to obtain
\begin{align}
\tilde{m} = -\mu_0 - s_\mu\tilde{x}_0 + \frac{s_\mu^2}{\mu_0}\, \sum_{n=1}^\infty |\tilde{x}_n|^2 + \mathcal{O}\left(\frac{s_\mu^3}{\mu_0^3}\right)\,.
\end{align}
Expanding $\phi$ in powers of $s_\mu/\mu_0$, we get with Eqs.~\eqref{eq:mu_0} and \eqref{eq:sigma_mu}
\begin{align}
\phi
=&-\theta_0
- s_\theta\, \tilde{x}_0
+ \frac{1}{2} \left(
\tilde{x}_0^2 + \frac{1}{\sin^2\left(\tfrac{\theta_0}{2}\right)}
\sum_{n=1}^\infty |\tilde{x}_n|^2
\right)\tan \left(\tfrac{\theta_0}{2}\right) s_\theta^2
+ \mathcal{O}\left(\frac{s_\theta^2}{\theta_0^2}\right)\,.
\end{align}
Knowing the distribution of $\tilde{x}_n$ (Eq.~\eqref{eq:x_tilde_correlations}), we can average over all $\tilde{x}_n$ to compute mean value and variance of $\phi$. We obtain
\begin{align}
\mathrm{E}[\phi]
&= -\theta_0+\frac{s_\theta^2}{2\sin(\theta_0)}
- \frac{d}{L}\ \frac{s_\theta^2}{2\tan(\theta_0)}
+ \mathcal{O}\left(\frac{s_\theta^3}{\theta_0^3}\right) \,,
\label{eq:E0:Dirac:phi_mean}
\\
\mathrm{Var}(\phi)
&=\frac{d s_\theta^2}{L}
+ \mathcal{O} \left(\frac{s_\theta^3}{\theta_0^3}\right)\,.
\label{eq:E0:Dirac:phi_sigma}
\end{align}

\subsubsection{Scattering phase for broken law of reflection}

For $\mu_0=0$, where the disorder-broadened edge states overlap with the Fermi energy $E=0$, we have
$\boldsymbol{\tilde{\mu}} = s_\mu \boldsymbol{\tilde{x}}$, and hence with Eqs.~\eqref{eq:mu_n_tilde_expanded}, \eqref{eq:sigma_mu}
\begin{align}
\tilde{m} = \frac{s_\theta}{2} \chi\,, \quad \text{with} \quad
\chi =  \tilde{x}_\uparrow^\dagger \boldsymbol{\tilde{x}}^{-1}\tilde{x}_\uparrow - \tilde{x}_0 \,.
\label{eq:m_tilde}
\end{align}
We obtain
\begin{align}
\phi
= \text{atan2}\left(1-s_\theta^2\chi^2/4, s_\theta \chi \right) \,.
\label{eq:phi_phase_mu_0=0}
\end{align}
For small $s_\theta$ we have $\phi = s_\theta \chi + \mathcal{O}(s_\theta^3)$, hence the distribution of $\phi$ is directly linked to the distribution of $\chi$, which we will now further explore.

Due to Eq.~\eqref{eq:x_tilde_correlations}, the elements of
$\boldsymbol{\tilde{x}}$ decay away from the diagonal,
$\mathrm{E}[|\tilde{x}_n|^2] \sim \exp(-n^2/2n_0^2)$. In the limit $n_0
\rightarrow 0$, which corresponds to the limit $d/L\rightarrow \infty$,
\emph{i.e.}, completely correlated (constant) disorder, the matrix $\boldsymbol{\tilde{x}}$ will therefore be essentially diagonal. In 0th order we have $\boldsymbol{\tilde{x}} = \tilde{x}_0 \mathbbm{1}$, and therefore
\begin{align}
\chi = \frac{\sum_{n=1}^\infty |\tilde{x}_n|^2}{\tilde{x}_0}  - \tilde{x}_0 \,.
\label{eq:chi_expansion_n0=0}
\end{align}
Assuming the $\tilde{x}_n$ to still be approximately independent (although the approximation made in Eq.~\eqref{eq:x_tilde_correlations} does not hold in the limit $d/L\rightarrow\infty$), due to the central limit theorem the numerator and denominator are independent Gaussian distributed variables, with zero (or approximately zero) mean. As a result, the first term $\chi_1$ of Eq.~\eqref{eq:chi_expansion_n0=0} follows a Cauchy distribution
\begin{align}
f_{\chi_1}(x)=\frac{1}{\pi}\frac{\gamma}{x^2+\gamma^2}\,.
\end{align}
However, its scale parameter scales as $\gamma \sim \exp(-1/n_0^2)$, therefore in the limit $n_0 \rightarrow 0$ we remain with the second term $\chi_0$ of Eq.~\eqref{eq:chi_expansion_n0=0}, $\chi = \chi_0 = -\tilde{x}_0$. In the limit $d/L\rightarrow\infty$ the approximation of Eq.~\eqref{eq:x_tilde_correlations} does not hold for $\tilde{x}_0$; instead we find $\mathrm{Var} (\tilde{x}_0)=1$, such that $\phi$ is normally distributed with mean $0$ and variance $s_\theta^2$.

In fact, we are however interested in the distribution of $\phi$ in the opposite limit, $L/d \rightarrow \infty$. In this limit $\chi_0 = -\tilde{x}_0$ becomes small, $\mathrm{Var}(\tilde{x}_0) = 1/\sqrt{2\pi n_0^2} =  d/L$, whereas we find numerically that $\chi_1$ still follows a Cauchy distribution, with a scale parameter $\gamma$ that becomes independent of $L/d$ and can be evaluated numerically as $\gamma\approx 0.8\,s_\theta$.
Remarkably, the value of $\gamma/ s_\theta$ we obtain numerically
is not universal, but depends weakly on the original distribution of the disorder $\theta(x)$. If we choose these parameters to be not normally distributed but to follow any other distribution, $\chi$ is still Cauchy distributed, but the scale parameter $\gamma$ will also depend on the higher cumulants of the chosen disorder distribution.

Based on this distribution, we can evaluate mean value and variance of the scattering phase $\phi$.
Since $f_{\chi_1}$ is even and $\phi$ is an odd function of $\chi$, we directly see that
$
\mathrm{E}[\phi]=0
$.
Furthermore, we can numerically evaluate the integral in $\mathrm{E}[\phi^2]$ to obtain
\begin{align}
\mathrm{Var}(\phi) &\approx 2.2 s_\theta\,.
\label{eq:Dirac:phi_sigma_preserved}
\end{align}

\subsection{Computation of the scattering matrix from the tight-binding model}

\subsubsection{Tight-binding Hamiltonian}

\begin{figure}[h!]
\centering
\includegraphics[width=.7\textwidth]{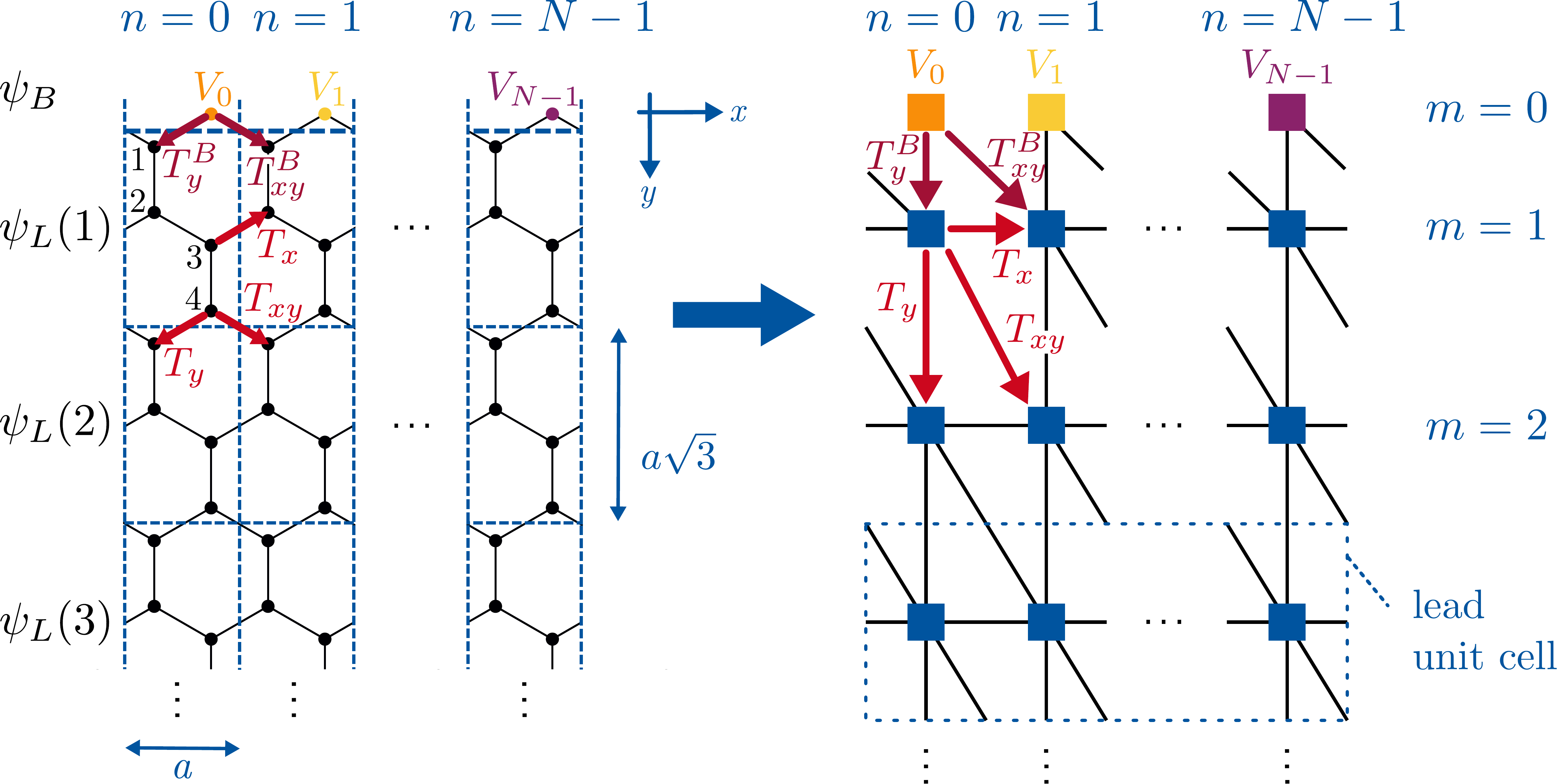}
\caption{Scheme of the tight-binding system for which the scattering matrix is computed, introducing a rectangular superlattice with four sites per unit cell. The superlattice unit cells are represented by blue squares in the right picture, defining a tight-binding system based on this superlattice.
$n$ labels the $x$-coordinate along the boundary from 0 to $N-1=(L-1)/a$ and $m$ is the corresponding $y$-index going from 1 to $\infty$ as the lead has translational invariance in $y$-direction. The index $m=0$ indicates the boundary sites that do not belong to the lead. Numbers in the (1,1) unit cell (left) specify the order of the sites within each unit cell in the vector representation that will be used. As indicated by half lines, periodic boundary conditions are implemented by connecting the $n=N-1$ to the $n=0$ cells. $V_n$ is the onsite disorder potential on the $n$-th boundary site and the $T$-s are the hopping matrices between adjacent superlattice unit cells within the lead and between lead and boundary (indicated by superscript $B$). $\psi_B$ is the wavefunction amplitude on the boundary sites and $\psi_L(m)$ the one on the $m$-th unit cell of the lead ($m$-th row of the superlattice).}
\label{fig:lattice}
\end{figure}

To extend the consideration to the more general case of two Dirac valleys, we compute the scattering matrix from an atomistic tight-binding model with nearest neighbor hopping $t\approx 2.8\,$eV, with a geometry as shown in Fig.~\ref{fig:lattice}, in direct analogy to Fig.~\ref{fig:geometry}. Disorder is included by introducing Gaussian distributed random on-site potentials $V_n$ on the boundary site $n$ with mean $V_d$ and variance $s_d^2$. The corresponding Hamiltonian is
\begin{align}
H = -t\sum_{\langle mnj,m^\prime n^\prime j^\prime\rangle}|m,n,j\rangle\langle m^\prime,n^\prime,j^\prime|
+ \sum_n V_n |0,n,4\rangle \langle 0,n,4 |
\label{eq:H_tb}
\end{align}
where the brackets under the first sum indicate that it goes only over nearest neighbors. According to Fig.~\ref{fig:lattice}, each lattice site is specified by three indices: $m$ labels the $y$-coordinate of the blue rectangular superlattice shown in Fig.~\ref{fig:lattice} from 1 to $\infty$ within the lead, being 0 on the boundary sites, $n$ labels the corresponding $x$-coordinate from 0 to $N-1=(L-1)/a$, with the boundary length $L$ and lattice constant $a$, and $j$ determines the position within each cell of the superlattice in the order specified in Fig.~\ref{fig:lattice}. The atomic orbitals $\{ |m,n,j \rangle |\}$ are assumed to form a complete basis of the lead Hilbert space within the tight-binding approximation.
Therefore, any state on the lead can be written as
\begin{align}
|\psi_L\rangle = \sum_{m=1}^\infty \sum_{n=0}^{N-1} \sum_{j=1}^4 \psi_L(m,n,j)|m,n,j\rangle\,,
\end{align}
where $\psi_L(m,n,j)=\langle m,n,j|\psi_L\rangle$ is the amplitude of the lead wavefunction on the lattice site $(m,n,j)$.
Correspondingly, the state on the boundary is given by the orbital states on the boundary sites as
\begin{align}
|\psi_B\rangle = \sum_{n=0}^{N-1} \psi_B(n)|0,n,4\rangle\,
\end{align}
with an amplitude $\psi_B(n)=\langle 0,n,4|\psi_B\rangle$ on the $n$-th boundary site.
We can write the combined wavefunction $\psi=\psi_L + \psi_B$ in a vector representation within this basis as
\begin{align}
\psi = \left(
\begin{matrix}
\psi_L \\
\psi_B
\end{matrix}
\right)
=\left(
\begin{matrix}
\vdots \\
\psi_L(2) \\
\psi_L(1) \\
\psi_B
\end{matrix}
\right)
=
\left(
\begin{matrix}
\vdots \\
\psi_L(2,N-1) \\
\vdots \\
\psi_L(2,0) \\
\psi_L(1,N-1) \\
\vdots \\
\psi_L(1,0) \\
\psi_B(N-1) \\
\vdots \\
\psi_B(0)
\end{matrix}
\right)
\,,\quad
\psi_L(m,n)
=
\left(
\begin{matrix}
\psi_L(m,n,1) \\
\psi_L(m,n,2) \\
\psi_L(m,n,3) \\
\psi_L(m,n,4)
\end{matrix}
\right)\,.
\label{eq:E0:tight_binding:general_wavefunction}
\end{align}
The tight-binding Hamiltonian Eq.~(\ref{eq:H_tb}) in matrix form reads
\begin{align}
H=\left(
\begin{matrix}
\ddots & \ddots & & \\
\ddots & H_L & T_L & \\
& T_L^\dagger & H_L & T_{LB} \\
& & T_{LB}^\dagger & H_B
\end{matrix}
\right)\,.
\label{eq:E0:tight_binding:Hamiltonian}
\end{align}
Here
\begin{align}
H_L = \left(
\begin{matrix}
H_0 & T_x & & T_x^\dagger \\
T_x^\dagger & H_0 & \ddots & \\
& \ddots & \ddots & T_x \\
T_x & & T_x^\dagger & H_0
\end{matrix}
\right)\,,
\quad
H_B = \left(
\begin{matrix}
V_{N-1} & & & \\
& \ddots & & \\
& & V_1 & \\
& & & V_0
\end{matrix}
\right)
\end{align}
are the $4N\times 4N$ Hamiltonian submatrix of each lead unit cell (row of the superlattice in Fig.~\ref{fig:lattice}, with fixed index $m\geq 1$) and the $N\times N$ submatrix of the boundary (containing the onsite disorder potential on the diagonal), respectively, and
\begin{align}
T_L=\left(
\begin{matrix}
T_y & T_{xy} & & \\
 & T_y & \ddots & \\
& & \ddots & T_{xy} \\
T_{xy} & & & T_y
\end{matrix}
\right)\,,\qquad
T_{LB}=\left(
\begin{matrix}
T_y^B & T_{xy}^B & & \\
 & T_y^B & \ddots & \\
& & \ddots & T_{xy}^B \\
T_{xy}^B & & & T_y^B
\end{matrix}
\right)
\end{align}
couple consecutive lead unit cells and the $m=1$ lead unit cell to the boundary, respectively.
Their corresponding subblocks are given in terms of the hopping parameter $t$ by
\begin{align}
H_0 = \left(
\begin{matrix}
 0 & -t &  0 &  0 \\
-t &  0 & -t &  0 \\
 0 & -t &  0 & -t \\
 0 &  0 & -t &  0
\end{matrix}
\right) \,,\quad
T_x = \left(
\begin{matrix}
0 & 0 & 0 & 0 \\
0 & 0 & -t & 0 \\
0 & 0 & 0 & 0 \\
0 & 0 & 0 & 0
\end{matrix}
\right) \,,
\quad
T_y = T_{xy} = \left(
\begin{matrix}
0 & 0 & 0 & -t \\
0 & 0 & 0 & 0 \\
0 & 0 & 0 & 0 \\
0 & 0 & 0 & 0
\end{matrix}
\right)\,,\quad
T_y^B = T_{xy}^B = \left(
\begin{matrix}
-t \\
0 \\
0 \\
0
\end{matrix}
\right) \,,
\end{align}
where $H_0$ contains all hoppings between sites within one superlattice unit
cell and the $T$'s the hoppings between adjacent cells, as sketched in
Fig.~\ref{fig:lattice}. For simplicity and to keep expressions shorter, we will
from now on set $t=1$, \emph{i.e.}, all energies such as the disorder potential will be given in units of $t$.

\subsubsection{Lead eigenstates}

To solve for the scattering matrix, we first have to compute propagating and evanescent eigenstates of the tight-binding Hamiltonian on an infinite lead without a boundary, which is given by
\begin{align}
H_L^\mathrm{inf}=\left(
\begin{matrix}
\ddots & \ddots & & \\
\ddots & H_L & T_L & \\
& T_L^\dagger & H_L & \ddots \\
& & \ddots & \ddots
\end{matrix}
\right)\,, \quad
\psi_L^\mathrm{inf} = \left(
\begin{matrix}
\vdots \\
\psi_L(2) \\
\psi_L(1) \\
\vdots
\end{matrix}
\right)\,.
\label{eq:appendix:S:infinite_lead}
\end{align}
Since this infinite lead has translational invariance in $x$- and in $y$-direction, we use a Bloch ansatz for the lead wavefunction,
\begin{align}
\psi_L^\mathrm{inf}(m,n) = \lambda^m\xi^n\chi\,,
\label{eq:appendix:S:Bloch_ansatz}
\end{align}
where $\lambda$ and $\xi$ are eigenvalues of the translation operator in $y$- and $x$-direction, respectively. The 4-vector $\chi$ gives the mode structure within each superlattice unit cell. Note that this Bloch ansatz lives on the rectangular superlattice. We thereby disregard the original honeycomb lattice structure and assume the hoppings $T_x$ and $T_y$ to be exactly aligned with the $x$- and $y$-axis, respectively, as shown in Fig.~\ref{fig:lattice} on the right. This means that we choose the mode structure $\chi$ within a unit cell to be multiplied by a factor of $\xi$ when hopping along $T_x$, by $\lambda$ when hopping along $T_y$, and by $\xi\lambda$ for hoppings $T_{xy}$. This choice amounts to a specific gauge of the phase of the wavefunction. Hence, it is completely equivalent to choosing Bloch phases according to the honeycomb structure by, \textit{e.g.}, assuming also a phase shift in $y$-direction for hoppings $T_x$.

As we have periodic boundary conditions $\psi_L^\mathrm{inf}(m,n)=\psi_L^\mathrm{inf}(m,n+N)$, it must hold that $\xi^N=1$, therefore we have
\begin{align}
\xi_\nu=e^{ik_{x,\nu}a}=e^{2\pi i\nu/N}
\quad \text{with} \quad
k_{x,\nu}=\frac{2\pi \nu}{L}=\frac{2\pi}{a}\frac{\nu}{N}\,,\quad
\nu=0,\dots,N-1\,.
\label{eq:appendix:S:pbc}
\end{align}
At $E_F=0$, the Fermi surface consists only of the Dirac points, so propagating modes have a momentum that lies at these points in momentum space. Therefore, the momentum $k_{x,\nu}$ must match the $x$-component of the Dirac points for some $\nu$ to have propagating modes at all, such that we demand $k_{x,\nu}=K_x=2\pi/3a$ or $k_{x,\nu}=K^\prime_x=-2\pi/3a$, since the Dirac points in this coordinate choice are given by $\boldsymbol{K} = 2\pi/3a (1,\sqrt{3})$, $\boldsymbol{K^\prime} = 2\pi/3a (-1,\sqrt{3})$. We conclude that $N\overset{!}{=}3\nu$, thus propagating modes are only possible if the boundary length $L$ is a multiple of $3a$, which we in the following will assume to be true.

Using the Bloch ansatz, the Schrödinger equation for the infinite lead at $E_F=0$ reduces to
\begin{align}
H_L^\mathrm{inf}\psi_L^\mathrm{inf}=0 \quad
\overset{\text{Eq.~(\ref{eq:appendix:S:Bloch_ansatz})}}{\Rightarrow}
\quad H_c \chi = 0 \,,
\end{align}
with
\begin{align}
H_c = H_0 + \left(\xi_\nu^{-1} T_x + \lambda^{-1} T_y + \xi_\nu^{-1}\lambda^{-1} T_{xy} + \text{h.c.} \right)\,.
\end{align}
By solving $\det H_c = 0$, we obtain the relations
\begin{align}
\lambda_\nu^+=\frac{(1+\xi_\nu)^2}{\xi_\nu}\,,\quad \lambda_\nu^-=\frac{\xi_\nu}{(1+\xi_\nu)^2}
\label{eq:appendix:S:lambda}
\end{align}
between the translation operator eigenvalues $\xi$ and $\lambda$ that need to be fulfilled for $H_c$ to have a zero eigenvalue. Thereby for each possible $x$-momentum $k_{x,\nu}$ we have two solutions for the momentum in $y$-direction defined through
\begin{align}
\lambda_\nu = e^{ik_{y,\nu}a}\,.
\end{align}
As we in fact restrict the lattice to positive $y$, modes do not have to be normalizable for negative $y$. Therefore, we can allow for all $ \lambda$ with $|\lambda|\leq 1$, \textit{i.e.}, for plane waves and modes that decay exponentially for $y\rightarrow \infty$.

We get the corresponding eigenmodes as
\begin{align}
\chi_\nu^+=\left(
\begin{matrix}
0 \\ -1 \\ 0 \\ 1+\xi_\nu
\end{matrix}
\right)\,,\quad
\chi_\nu^-=\left(
\begin{matrix}
-1-\xi_\nu \\ 0 \\ \xi_\nu \\ 0
\end{matrix}
\right)\,.
\label{eq:appendix:S:chi}
\end{align}
Regarding their translation eigenvalue $\lambda_\nu^\pm$, these lead eigenmodes can be classified as follows:

\paragraph{Propagating modes:}

Modes with $\xi_{\boldsymbol{K}}=\xi_{N/3}=e^{2\pi i/3}$ and $\xi_{\boldsymbol{K^\prime}}=\xi_{2N/3}=e^{-2\pi i/3}$ have $\lambda=1$ and thus are propagating (their amplitudes do not decay in $y$-direction). Defining
\begin{align}
\Phi_\mathrm{pr}=(\chi_{\boldsymbol{K}}^-,\,\chi_{\boldsymbol{K}}^+,\,\chi_{\boldsymbol{K^\prime}}^-,\,\chi_{\boldsymbol{K^\prime}}^+)\,,\quad
\Xi_\mathrm{pr}=\text{diag}(\xi_{\boldsymbol{K}},\,\xi_{\boldsymbol{K}},\,\xi_{\boldsymbol{K^\prime}},\,\xi_{\boldsymbol{K^\prime}})\,,\quad
\Lambda_\mathrm{pr} = \mathbbm{1}_4\,,
\label{eq:appendix:S:Phi_p}
\end{align}
the set of propagating modes on the $m$-th lead unit cell is given by
\begin{align}
\Psi_\mathrm{pr}\Lambda_\mathrm{pr}^m=\left(
\begin{matrix}
\Phi_\mathrm{pr} \,\Xi_\mathrm{pr}^{N-1} \\
\vdots \\
\Phi_\mathrm{pr} \,\Xi_\mathrm{pr} \\
\Phi_\mathrm{pr}
\end{matrix}
\right)\Lambda_\mathrm{pr}^m\,.
\end{align}
To separate incoming and outgoing states, we have to find eigenstates of the particle current operator
\begin{align}
J=\frac{2a}{\hbar}\mathfrak{Im}(\Lambda_\mathrm{pr}^*T_L)
\end{align}
within the set of propagating modes. Therefore we have to diagonalize
\begin{align}
J_\mathrm{pr} = \Psi_\mathrm{pr}^\dagger J \Psi_\mathrm{pr}
= \frac{2 a}{\hbar} \, \Psi_\mathrm{pr}^\dagger \,\mathfrak{Im}(\Lambda_\mathrm{pr}^* T_L)\,\Psi_\mathrm{pr}
= \frac{a}{i\hbar}\, \Psi_\mathrm{pr}^\dagger (T_L-T_L^\dagger) \Psi_\mathrm{pr}\,.
\end{align}
$J_\mathrm{pr}$ can be straightforwardly evaluated from the definitions above, yielding
\begin{align}
J = \frac{4}{\sqrt{3}}\,Nv_F
 \left(
\begin{matrix}
0 & -\alpha & & & \\
-\alpha^* & 0 & & \\
& & 0 & \alpha^* \\
& & \alpha & 0
\end{matrix}
\right)\,,
\end{align}
with the Fermi velocity $v_F = \sqrt{3}ta/2\hbar$ and $\alpha = \xi_{\boldsymbol{K}}^{1/4} = e^{i\pi/6}\,$.
The eigenvalues of J are $-\frac{4}{\sqrt{3}} \,Nv_F$ (corresponding to incoming modes) with normalized eigenvectors $v^-_{\boldsymbol{K}}=\frac{1}{\sqrt{2}}(\alpha,1,0,0)^T$ and $v^-_{\boldsymbol{K^\prime}}=\frac{1}{\sqrt{2}}(0,0,-\alpha^*,1)^T$, and $+\frac{4}{\sqrt{3}} \,Nv_F$ (corresponding to outgoing modes) with eigenvectors $v^+_{\boldsymbol{K}}=\frac{1}{\sqrt{2}}(-\alpha,1,0,0)^T$ and $v^+_{\boldsymbol{K^\prime}}=\frac{1}{\sqrt{2}}(0,0,\alpha^*,1)^T$. Since we sorted the propagating modes by the two valleys (Eq.~(\ref{eq:appendix:S:Phi_p})) and these eigenvectors do not mix the subspaces of the two valleys, we can also assign each to a unique valley by labeling them $\boldsymbol{K},\boldsymbol{K^\prime}$.

To ensure that $S$ is unitary, all propagating lead eigenmodes have to be properly normalized to carry the same probability current. However, as here all modes have already the same current eigenvalue according to its absolute value, we can choose any normalization that simplifies the calculation.
With
\begin{align}
\Phi_\mathrm{in}&=\Phi_\mathrm{pr}\cdot(v^-_{\boldsymbol{K}},\,v^-_{\boldsymbol{K^\prime}})/\sqrt{N}\,,\quad
\Xi_\mathrm{in}=\text{diag}(\xi_{\boldsymbol{K}},\,\xi_{\boldsymbol{K^\prime}})\,,\quad
\Lambda_\mathrm{in}=\mathbbm{1}_2 \,,
\quad \text{and}\\
\Phi_\mathrm{out}&=\Phi_\mathrm{pr}\cdot(v^+_{\boldsymbol{K^\prime}},\,v^+_{\boldsymbol{K}})/\sqrt{N}\,,\quad
\Xi_\mathrm{out}=\text{diag}(\xi_{\boldsymbol{K^\prime}},\,\xi_{\boldsymbol{K}})\,,\quad
\Lambda_\mathrm{out}=\mathbbm{1}_2 \,,
\end{align}
we can therefore define incoming and outgoing modes with current normalized to $\mp 4v_F/\sqrt{3}$ and well-defined momenta $k_x= 2\pi/3,k_x=-2\pi/3$ on the $m$-th lead unit cell within the notation introduced in Eq.~(\ref{eq:E0:tight_binding:general_wavefunction}) as
\begin{align}
(\psi^\mathrm{in}_{\boldsymbol{K}}(m),\psi^\mathrm{in}_{\boldsymbol{K^\prime}}(m))
&=\Psi_\mathrm{in} \Lambda_\mathrm{in}^m =\left(
\begin{matrix}
\Phi_\mathrm{in} \,\Xi_\mathrm{in}^{N-1} \\
\vdots \\
\Phi_\mathrm{in} \,\Xi_\mathrm{in} \\
\Phi_\mathrm{in}
\end{matrix}
\right) \Lambda_\mathrm{in}^m
\,,\\
(\psi^\mathrm{out}_{\boldsymbol{K^\prime}}(m),\psi^\mathrm{out}_{\boldsymbol{K}}(m))
&=\Psi_\mathrm{out} \Lambda_\mathrm{out}^m =\left(
\begin{matrix}
\Phi_\mathrm{out} \,\Xi_\mathrm{out}^{N-1} \\
\vdots \\
\Phi_\mathrm{out} \,\Xi_\mathrm{out} \\
\Phi_\mathrm{out}
\end{matrix}
\right) \Lambda_\mathrm{out}^m \,.
\end{align}
Note that we sort the outgoing modes in opposite order with respect to the valleys as the incoming modes. This is to ensure that they reflect time-reversal symmetry. Under time-reversal the velocity of the modes is reversed and the valleys are exchanged. Therefore, with this ordering the outgoing modes are the time-reversed incoming ones.

\paragraph{Evanescent modes:}

For $-N/3<\nu<N/3$ 
holds $\lambda_\nu^-<1$ and $\lambda_\nu^+>1$, thus $\chi_\nu^-$-modes are evanescent, whereas $\chi^+$-modes are not normalizable.
For $N/3<\nu<2N/3=-N/3\mod N$ 
the opposite case is true. The normalization of the evanescent modes is irrelevant for the result of the calculation of $S$, therefore we multiply them with $1/\sqrt{N}$ which will later simplify prefactors. We can then simply write the set of evanescent modes within the $m$-th lead unit cell as
\begin{align}
\Big(\psi^\mathrm{ev}_{-\frac{N}{3}+1}(m), \dots,
\psi^\mathrm{ev}_{\frac{N}{3}-1}(m),
\psi^\mathrm{ev}_{\frac{N}{3}+1}(m) &, \dots,
\psi^\mathrm{ev}_{\frac{2N}{3}-1}(m)\Big)
=\Psi_\mathrm{ev}\Lambda_\mathrm{ev}
= (\Psi_\mathrm{ev}^-\Lambda_\mathrm{ev}^-,
\Psi_\mathrm{ev}^+\Lambda_\mathrm{ev}^+)
\,,
\end{align}
where
\begin{align}
\Psi_\mathrm{ev}^\mp=
\frac{1}{\sqrt{N}}
\left(
\begin{matrix}
\Phi_\mathrm{ev}^\mp \,(\Xi_\mathrm{ev}^\mp)^{N-1} \\
\vdots \\
\Phi_\mathrm{ev}^\mp \,\Xi_\mathrm{ev}^\mp \\
\Phi_\mathrm{ev}^\mp
\end{matrix}
\right)\,,\quad
\Lambda_\mathrm{ev} = \text{diag}(\Lambda_\mathrm{ev}^-,\Lambda_\mathrm{ev}^+)\,,
\end{align}
with
\begin{align}
\Phi_\mathrm{ev}^-&=(\chi^-_{-\frac{N}{3}+1}, \dots, \,\chi^-_{\frac{N}{3}-1})\,,\quad
\Xi_\mathrm{ev}^-=\text{diag}(\xi_{-\frac{N}{3}+1}, \dots, \,\xi_{\frac{N}{3}-1})\,,\quad
\Lambda_\mathrm{ev}^-=\text{diag}(\lambda^-_{-\frac{N}{3}+1}, \dots, \,\lambda^-_{\frac{N}{3}-1})\,,
\nonumber \\
\Phi_\mathrm{ev}^+&=(\chi^+_{\frac{N}{3}+1}, \dots, \,\chi^+_{\frac{2N}{3}-1})\,,\quad
\Xi_\mathrm{ev}^+=\text{diag}(\xi_{\frac{N}{3}+1}, \dots, \,\xi_{\frac{2N}{3}-1})\,,\quad
\Lambda_\mathrm{ev}^+=\text{diag}(\lambda^+_{\frac{N}{3}+1}, \dots, \,\lambda^+_{\frac{2N}{3}-1})\,.
\end{align}

\subsubsection{Computation of the scattering matrix}

Since we have two incoming and two outgoing modes at the Dirac points, the scattering matrix is 2 x 2. It can be parametrized as
\begin{align}
S = e^{i\phi} \left(
\begin{matrix}
re^{i\Delta} & \sqrt{1-r^2} \\ \sqrt{1-r^2} & -re^{-i\Delta}
\end{matrix} \right) \,,
\end{align}
with three real parameters $r$, $\phi$ and $\Delta$. The phase $\phi$ of the off-diagonal (intra-valley) elements is the direct analogue of the scattering phase within the single-valley continuum description.

To solve for the scattering matrix, we use the eigenstates of the infinite lead to compose scattering states in the lead, now assuming to have the boundary terminating the lead, which are given in the $m$-th lead unit cell by
\begin{align}
\Psi_L(m) =
(\psi_{\boldsymbol{K}}(m),\psi_{\boldsymbol{K^\prime}}(m))
= \Psi_\mathrm{in} \Lambda_\mathrm{in}^m + \Psi_\mathrm{out} \Lambda_\mathrm{out}^m S + \Psi_\mathrm{ev}\Lambda_\mathrm{ev}^m S_\mathrm{ev}\,,
\end{align}
Each of these two scattering states is a superposition of a fixed incoming mode with momentum $\boldsymbol{K}$ or $\boldsymbol{K^\prime}$, outgoing modes into which the incoming mode has been reflected at the boundary (expressed by the scattering matrix $S$), and evanescent modes, where $S_\mathrm{ev}$ gives the amplitudes to scatter into them, equivalently to $S$.
With $\Psi_B=(\psi_{B,\boldsymbol{K}},\psi_{B,\boldsymbol{K^\prime}})$, where the additional subscript $\boldsymbol{K},\boldsymbol{K^\prime}$ distinguishes the boundary wavefunctions depending on the momentum of the incoming modes, the last two blocks of the Schrödinger equation
$
H\left(\begin{matrix}
\Psi_L \\ \Psi_B
\end{matrix}\right)
=0
$
yield
\begin{align}
\left(
\begin{matrix}
T_L \Psi_\mathrm{out} & T_L \Psi_\mathrm{ev} & -T_{LB} \\
T_{LB}^\dagger \Psi_\mathrm{out} \Lambda_\mathrm{out} & T_{LB}^\dagger \Psi_\mathrm{ev} \Lambda_\mathrm{ev} & H_B
\end{matrix}
\right)
\left(
\begin{matrix}
S \\ S_\mathrm{ev} \\ \Psi_B
\end{matrix}
\right)
=
\left(
\begin{matrix}
-T_L \Psi_\mathrm{in} \\ -T_{LB}^\dagger \Psi_\mathrm{in} \Lambda_\mathrm{in}
\end{matrix}
\right)\,.
\label{eq:appendix:S:original_system}
\end{align}
We find that $T_L \Psi_\mathrm{ev}^-=0$, $T_{LB}^\dagger \Psi_\mathrm{ev}^+=0$.
Applying a discrete Fourier transform of both equation blocks by multiplying from the left by
\begin{align}
\left(
\begin{matrix}
U\otimes \mathbbm{1}_4 & 0 \\
0 & U
\end{matrix}
\right)\,, \quad
U_{mn} = \frac{1}{\sqrt{N}}\, e^{2\pi i (N-m)(N-n)/N}
= \frac{1}{\sqrt{N}}\, e^{2\pi i mn/N}\,,
\label{eq:appendix:S:DFT}
\end{align}
and explicitly computing all blocks of the system using the definitions given before, we get
\begin{align}
\left(
\renewcommand{\arraystretch}{1.3}
\begin{array}{c c c c | c}
		0 & 0 & 0 & 0 & \\
		-A_2 & 0 & 0 & 0 & \\
		0 & 0 & 0 & \Lambda_\mathrm{ev}^+ & \mathbbm{1} + \Xi^\dagger \\
		-A_1 & 0 & 0 & 0 & \\
		0 & 0 & 0 & 0 & \\ \hline
		0 & 0 & \Xi_\mathrm{ev}^{->} & 0 & \\
		\alpha^*A_2 & 0 & 0 & 0 & \\
		0 & 0 & 0 & 0 & U H_B U^\dagger \\
		-\alpha A_1 & 0 & 0 & 0 & \\
		0 & \Xi_\mathrm{ev}^{-<} & 0 & 0 &
\end{array}
\right)
\left(
\begin{matrix}
S \\ S_\mathrm{ev}^{-<} \\ S_\mathrm{ev}^{->} \\ S_\mathrm{ev}^+ \\ U\Psi_B
\end{matrix}
\right)
=
\left(
\renewcommand{\arraystretch}{1.3}
\begin{array}{c}
0 \\ A_1 \\ 0 \\ A_2 \\ 0 \\ \hline
0 \\ \alpha^*A_1 \\ 0 \\ -\alpha A_2 \\ 0
\end{array}
\right)
\label{eq:appendix:S:complete_system}
\end{align}
with
\begin{align}
A_1 = \left(\frac{1}{\sqrt{2}}, 0\right)\,,\quad A_2 = \left(0, \frac{1}{\sqrt{2}}\right)\,,
\end{align}
and
\begin{align}
\Xi_\mathrm{ev}^{-<}&=\text{diag}(\xi_{-\frac{N}{3}+1}, \dots, \,\xi_0)\,, \quad
\Xi_\mathrm{ev}^{->}=\text{diag}(\xi_{1}, \dots, \,\xi_{\frac{N}{3}-1})\,, \quad
\Xi_\mathrm{ev}^-=\text{diag}(\Xi_\mathrm{ev}^{-<},\Xi_\mathrm{ev}^{->})\,,
\nonumber \\
\Xi&=\text{diag}(\xi_1,\dots,\,\xi_N)
=\text{diag}(\Xi_\mathrm{ev}^{->}, \xi_{\boldsymbol{K}}, \Xi_\mathrm{ev}^+, \xi_{\boldsymbol{K^\prime}}, \Xi_\mathrm{ev}^{-<})\,.
\end{align}
Correspondingly, the evanescent modes scattering matrix $S_\mathrm{ev}$ is split up into parts for the same momentum ranges as
$
S_\mathrm{ev}=\left(
S_\mathrm{ev}^{-<} , S_\mathrm{ev}^{->} , S_\mathrm{ev}^+
\right)^T.
$
The lower right block of Eq.~(\ref{eq:appendix:S:complete_system}) has the form
\begin{align}
U H_B U^\dagger
=
\frac{1}{\sqrt{N}}
\left(
\begin{matrix}
\tilde{V}_0 & \tilde{V}_1 & \hdots & \tilde{V}_{N-1} \\
\tilde{V}_1^* & \tilde{V}_0 & \ddots & \vdots \\
\vdots & \ddots & \ddots & \tilde{V}_1 \\
\tilde{V}_{N-1}^* & \hdots & \tilde{V}_1^* & \tilde{V_0}
\end{matrix}
\right)\,,
\end{align}
with the Fourier coefficients of the disorder potential
\begin{align}
\tilde{V}_k
=\frac{1}{\sqrt{N}}\sum\limits_{j=0}^{N-1} V_j\, e^{2\pi i (j-1) k /N}\,.
\label{eq:appendix:S:V_k_tilde}
\end{align}
By clever pivoting, \textit{i.e.}, exchanging the rows and columns of Eq.~(\ref{eq:appendix:S:complete_system}), we can bring the system
into a block-diagonal form where the lower right $((N/3+3) \times (N/3+3))$-block does not depend on $S_\mathrm{ev}$, thus leaving us with
\begin{align}
\left(
\begin{matrix}
\bar{V} & B_1 \\ C & D_1
\end{matrix}
\right)
\left(
\begin{matrix}
\bar{\Psi} \\ S
\end{matrix}
\right)
=
\left(
\begin{matrix}
B_2 \\ D_2
\end{matrix}
\right)\,.
\label{eq:appendix:S:reduced_system}
\end{align}
Here $\bar{\Psi}$ contains some of the components of $U\Psi_B$ which however will be eliminated in the procedure of solving for $S$ and therefore do not need to be specified. The matrix
\begin{align}
\bar{V}=
\frac{1}{\sqrt{N}}
\left(
\renewcommand{\arraystretch}{2}
\begin{array}{c c c|c c}
\tilde{V}_0 & \hdots & \tilde{V}_{\frac{N}{3}-2} & \tilde{V}_1^* & \tilde{V}_{\frac{N}{3}-1}\\
\vdots & \ddots & \vdots & \vdots & \vdots \\
\tilde{V}_{\frac{N}{3}-2}^* & \hdots & \tilde{V}_0 & \tilde{V}_{\frac{N}{3}-1}^* & \tilde{V}_1 \\ \hline
\tilde{V}_1 & \hdots & \tilde{V}_{\frac{N}{3}-1} & \tilde{V}_0 & \tilde{V}_{\frac{N}{3}} \\
\tilde{V}_{\frac{N}{3}-1}^* & \hdots & \tilde{V}_1^* & \tilde{V}_{\frac{N}{3}}^* & \tilde{V}_0
\end{array}
\right)
\end{align}
contains only the lowest third of the Fourier components of the disorder potential $V_j$.
Further, the remaining blocks of the system are given by
\begin{align}
C &=
\left(
\begin{array}{c c c|c c}
0 & \hdots & 0 & 1+\xi_{\boldsymbol{K^\prime}} & 0 \\
0 & \hdots & 0 & 0 & 1+ \xi_{\boldsymbol{K}}
\end{array}
\right)\,,
\nonumber
\\
B_1 &=
\left(
\renewcommand{\arraystretch}{1.3}
\begin{array}{c c}
0 & 0 \\
\vdots & \vdots \\
0 & 0 \\ \hline
0 & \frac{\alpha^*}{\sqrt{2}} \\
\frac{-\alpha}{\sqrt{2}} & 0
\end{array}
\right)\,,\quad
B_2 =
\left(
\renewcommand{\arraystretch}{1.3}
\begin{array}{c c}
0 & 0 \\
\vdots & \vdots \\
0 & 0 \\ \hline
\frac{\alpha^*}{\sqrt{2}} & 0 \\
0 & \frac{-\alpha}{\sqrt{2}}
\end{array}
\right)\,,
\quad
D_1 =
\left(
\begin{array}{c c}
0 & -\frac{1}{\sqrt{2}} \\
-\frac{1}{\sqrt{2}} & 0
\end{array}
\right)\,,\quad
D_2 =
\left(
\begin{array}{c c}
\frac{1}{\sqrt{2}} & 0 \\
0 & \frac{1}{\sqrt{2}}
\end{array}
\right)\,.
\end{align}
Assuming the invertibility of $\bar{V}$ and $D_1-C\bar{V}^{-1}B_1$, we can use standard block matrix inversion to solve Eq.~(\ref{eq:appendix:S:reduced_system}) by multiplying with
$\left(\begin{matrix}
\bar{V} & B_1 \\ C & D_1
\end{matrix}\right)^{-1}$
from the left.
We thereby obtain
\begin{align}
S = \left[D_1-C\bar{V}^{-1}B_1\right]^{-1}(D_2 - C\bar{V}^{-1}B_2)\,,
\label{eq:appendix:S:S(Vbar)}
\end{align}
reducing the problem to the inversion of $\bar{V}$.
Due to the structure of $B_1$, $B_2$, and $C$, we only need to know the lower right $2\times 2$ block of $\bar{V}^{-1}$, which we denote by
\begin{align}
\bar{V}^{-1}=
\left(
\begin{array}{c c|c c}
\ddots & & \ddots & \\
& \ddots & & \ddots \\ \hline
\ddots & & W_{11} & W_{12} \\
& \ddots & W_{21} & W_{22}
\end{array}
\right)\,.
\end{align}
As $\bar{V}$ (and therefore also $W$) is Hermitian, it must hold that $W_{11}, W_{22}\in \mathbb{R}$ and $W_{21}=W_{12}^*$. Further, from the structure of $\bar{V}$ we conclude that $W_{11}=W_{22}$. We formally obtain $Y=W^{-1}$ by again using block matrix inversion. $Y$ is then given by the Schur complement of the upper left block of $\bar{V}$ as
\begin{align}
Y &= \left(
\begin{matrix}
Y_{11} & Y_{12} \\
Y_{12}^* & Y_{11}
\end{matrix}
\right)
=
\frac{1}{\sqrt{N}}
\left[
\left(
\renewcommand{\arraystretch}{1.3}
\begin{array}{c c}
\tilde{V}_0 & \tilde{V}_{\frac{N}{3}} \\
\tilde{V}_\frac{N}{3}^* & \tilde{V}_0
\end{array}
\right)
-
\left(
\renewcommand{\arraystretch}{1.3}
\begin{array}{c c c}
\tilde{V}_1 & \hdots & \tilde{V}_{\frac{N}{3}-1} \\
\tilde{V}_{\frac{N}{3}-1}^* & \hdots & \tilde{V}_1^*
\end{array}
\right)
\left(
\begin{matrix}
\tilde{V}_0 & \hdots & \tilde{V}_{\frac{N}{3}-2} \\
\vdots & \ddots & \vdots \\
\tilde{V}_{\frac{N}{3}-2}^* & \hdots & \tilde{V}_0
\end{matrix}
\right)^{-1}
\left(
\begin{matrix}
\tilde{V}_1^* & \tilde{V}_{\frac{N}{3}-1} \\
\vdots & \vdots \\
\tilde{V}_{\frac{N}{3}-1}^* & \tilde{V}_1 \\
\end{matrix}
\right)
\right]
\,.
\label{eq:appendix:S:Y}
\end{align}
From Eq.~(\ref{eq:appendix:S:S(Vbar)}) we can straightforwardly write down S in terms of the $Y_{ij}$, resulting in
\begin{align}
S = \frac{1}{1-\det Y +2iY_{11}}
\left(
\begin{matrix}
2\alpha Y_{12}^* & 1+\det Y \\
1+\det Y & -2\alpha^* Y_{12}
\end{matrix}
\right)\,.
\label{eq:E0:tight_binding:S_result}
\end{align}
The scattering phase $\phi$ can be obtained as
\begin{align}
\phi &= \arg(S_{12}) = \mathrm{atan2}\big(1-\det Y, -2Y_{11}\big)\,.
\label{eq:E0:tight_binding:phi_general}
\end{align}

\subsubsection{Distribution of the scattering phase}

Since the structure of the $Y_{ij}$ is completely analogous to that of $\tilde{m}$ in Eq.~\eqref{eq:phi_Dirac}, the same reasoning can be applied to distinguish whether or not the boundary overlaps with the band of edge states.
When the edge states are shifted away from $E=0$, \emph{i.e.}, in the limit $s_d \ll V_d$, we obtain
\begin{align}
\mathrm{E}[\phi] &= -\arctan\left( \frac{2V_d}{1-V_d^2} \right)
+ \frac{2}{3}\frac{s_d^2}{V_d\left(1+V_d^2\right)}
- \frac{a}{L}\cdot\frac{2\left(1-V_d^2 \right)}{V_d\left(1+V_d^2\right)^2} \ s_d^2
+ \mathcal{O}\left(\frac{s_d^3}{V_d^3}\right)\,,
\label{eq:E0:tight_binding:phi_mean}
\\
\mathrm{Var}(\phi) &= \frac{a}{L} \frac{4 s_d^2}{\left(1+V_d^2\right)^2}
+ \mathcal{O}\left(\frac{s_d^3}{V_d^3}\right)\,.
\label{eq:E0:tight_binding:phi_sigma}
\end{align}
For a boundary with $V_d=0$ that fulfills the condition of diffusive scattering, we can again not solve for $\phi$ as a function of $L$. We can however compute an explicit expression for $L=6$,
\begin{align}
\mathrm{E} [\phi_6 ] = 0\,, \quad
\mathrm{Var}(\phi_6) = \sqrt{\frac{4\pi}{3}}\,\ln(2)\, s_d + \mathcal{O}\left(s_d^2\right)\approx 1.4 \, s_d \,,
\label{eq:E0:tight_binding:N=6}
\end{align}
which we find numerically to be in good agreement with the results for larger $L$.

\subsection{Comparison between Dirac equation and tight-binding model}

The results obtained from the Dirac equation and the tight-binding model show qualitative agreement for both broken and preserved law of reflection. To also find a quantitative relation, we compare Eq.~\eqref{eq:E0:Dirac:phi_sigma} to Eq.~\eqref{eq:E0:tight_binding:phi_sigma} and Eq.~\eqref{eq:Dirac:phi_sigma_preserved} to Eq.~\eqref{eq:E0:tight_binding:N=6}. To equate the results of the two models for both cases, we have to assume that the scattering phases in the two models are not the same (due to the different number of modes), but related by some factor, $\mathrm{Var}_{\text{tb}}(\phi) \propto \mathrm{Var}_\text{Dirac}(\phi)$. Also assuming $s_d \propto s_\theta$, we find $\mathrm{Var}_{\text{tb}}(\phi) \simeq 0.1 \, \mathrm{Var}_\text{Dirac}(\phi)$ and $s_d \simeq 0.1\, s_\theta$.

\subsection{Magnetic focusing conductance in the absence of edge disorder}
\begin{figure}[!tb]
\includegraphics[width=0.5\columnwidth]{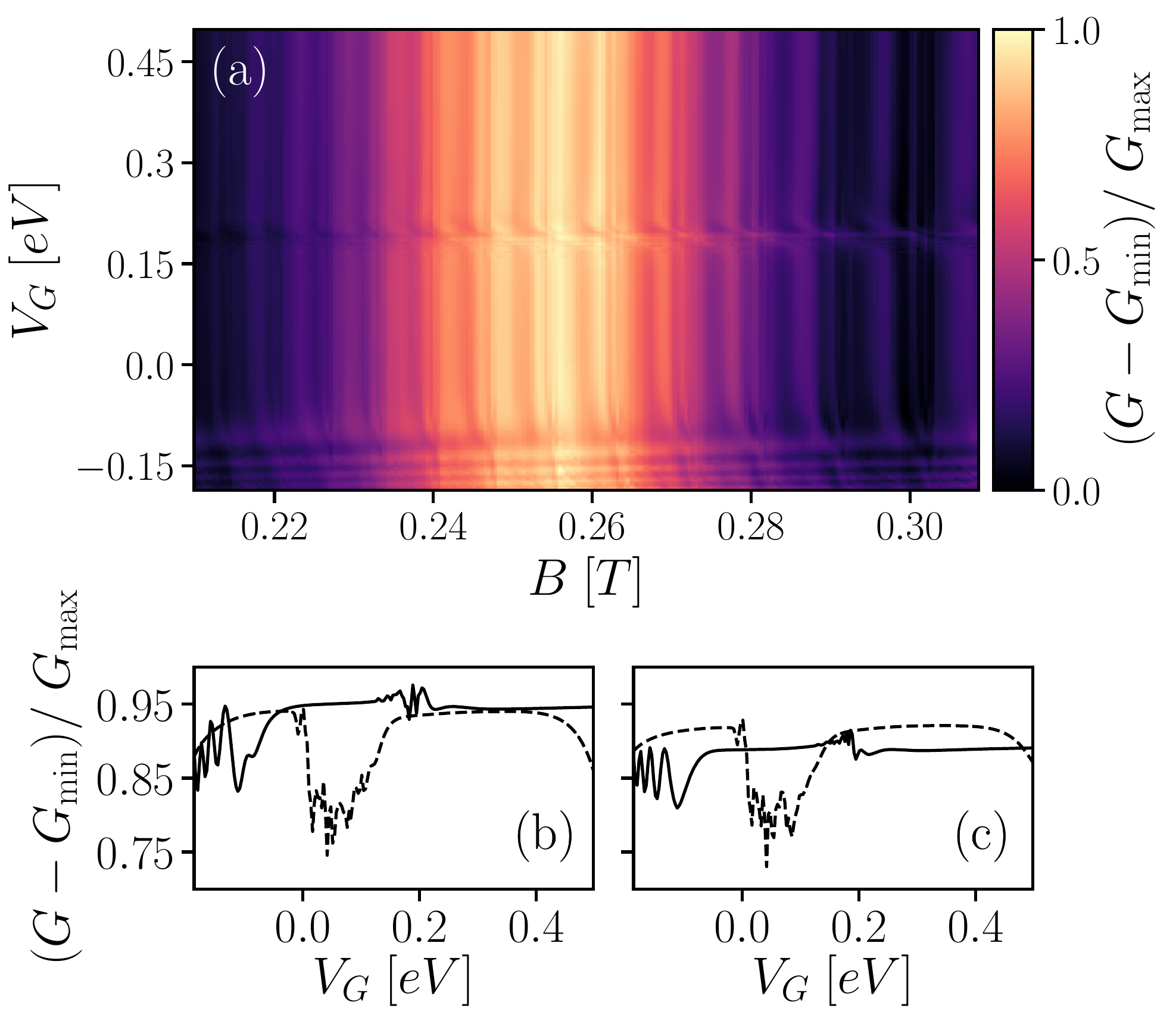}
\caption{
(a) Conductance around the second focusing peak at $E_F = 0.093$ eV versus gate
voltage without edge disorder, \emph{i.e.}, $V_d = s_d = 0$.
(b) Line cuts of the focusing conductance at $B=0.255\,$T versus gate voltage, without edge disorder (solid line) from (a), and with edge disorder from Fig.~3 (b) of the main text (dashed line).
(c) Line cuts of the focusing conductance versus gate voltage averaged over the magnetic field values $0.247 \leq B \leq 0.259$ T at the second focusing resonance, without (solid line) and with (dashed line) edge disorder.
The data with disorder is taken from Fig.~3 (b) of the main text.
In the absence of edge disorder, a small region of resonant conductance peaks appears around $V_G = 0.2$ eV, as the average potential in the boundary aligns with the Fermi level, but unlike the case with disorder, no clear dip is present.
When the boundary is clean, the gate forms a quantum well by the boundary at large negative gate potentials $V_G \lesssim -0.15$ eV, resulting in resonant oscillations in the conductance.
Similar oscillations also appear in the case with edge disorder, but at even larger negative gate potentials because of an overall average potential shift by the boundary due to onsite disorder, $V_d = 0.062$ eV.
}
\label{fig:focus_no_dis}
\end{figure}
In order to verify that the conductance dip at the second focusing resonance is a consequence of edge disorder, we compare the focusing conductance with edge disorder to the focusing conductance of a device with a clean boundary.
The setup is otherwise the same as the one we present in the main text, with the parameters identical to those used to obtain Figs.~3 (b)-(d) of the main text.

The results are shown in Fig.~\ref{fig:focus_no_dis}, where
Fig.~\ref{fig:focus_no_dis} (a) shows the focusing conductance versus gate
voltage and magnetic field strength without disorder, \emph{i.e.}, with $V_d = s_d = 0$.
The average potential at the boundary aligns with the Fermi level at a gate voltage $V_G \approx 0.2$ eV, which is larger than in the case with disorder included (see also Fig.~3 (b) of the main text).
This distinction arises due to the difference in the average edge disorder potential $V_d$, which is nonzero when disorder is included.
At large negative gate potentials $V_G \lesssim -0.15$ eV, resonant conductance oscillations also appear because the gate forms a quantum well by the boundary.
A similar phenomenon occurs in the case with edge disorder, but outside the energy window we consider, and is unrelated to the mechanism we are investigating.
In Fig.~\ref{fig:focus_no_dis} (a), some conductance oscillations appear in the conductance around the charge neutrality point $V_G \approx 0.2$ eV, but no clear dip is visible.
Furthermore, Fig.~\ref{fig:focus_no_dis} (b) gives a comparison of conductance line cuts at $B = 0.255$ T for a clean boundary with the case including edge disorder from Fig.~3 (b) of the main text.
We see that the oscillations when the edge potential aligns with the Fermi level in the clean case are much smaller in scale than the conductance dip that appears with the inclusion of edge disorder.
The same trends are visible in Fig.~\ref{fig:focus_no_dis} (c), which compares the focusing conductance averaged over magnetic field values at the second focusing peak, with and without edge disorder.
Therefore, we conclude that the dip in the focusing conductance at the second focusing peak arises due to edge disorder, namely when the average potential at the boundary aligns the disordered band of edge states with the Fermi level.

\end{document}